\documentclass[pra,onecolumn,floatfix,a4paper,superscriptaddress]{revtex4}
\usepackage{bm,color,graphicx,amsmath,txfonts}
\usepackage{here}
\usepackage{array}
\usepackage{graphicx}
\usepackage[colorlinks, citecolor=blue,linkcolor=blue]{hyperref}
\usepackage{braket}
\usepackage{dsfont}
\newcommand{\e}{{e}}

\begin{document}

\makeatletter
\renewcommand{\@biblabel}[1]{\makebox[2em][l]{\textsuperscript{\textcolor{black}{\fontsize{10}{12}\selectfont[#1]}}}}
\makeatother

\let\oldbibliography\thebibliography
\renewcommand{\thebibliography}[1]{%
  \addcontentsline{toc}{section}{\refname}%
  \oldbibliography{#1}%
  \setlength\itemsep{0pt}%
}

\title{Controlling the Dynamical Evolution of Quantum Coherence and Quantum Correlations in $ e^{+}e^{-} \to \Lambda\bar{\Lambda}$ Processes at BESIII}

\author{Elhabib Jaloum}
\affiliation{LPTHE-Department of Physics, Faculty of Sciences, Ibnou Zohr University, Agadir, Morocco}

\author{Mohamed Amazioug}
\email{m.amazioug@uiz.ac.ma}
\affiliation{LPTHE-Department of Physics, Faculty of Sciences, Ibnou Zohr University, Agadir, Morocco}

\date{\today}

\begin{abstract}

Quantum coherence, a cornerstone of quantum mechanics, is of paramount importance for quantum information protocols. However, maintaining coherence in elementary particle systems presents significant challenges. In this work, we investigate quantum coherence and quantum correlations in the $e^{+}e^{-} \to \Lambda\bar{\Lambda}$ processes at BESIII using experimentally feasible parameters, where $\Lambda$ and $\bar{\Lambda}$ denote the spin-$1/2$ hyperon and its antihyperon, respectively. We analyze the dependence of quantum coherence and quantum correlations on the scattering angle $\varphi$. Notably, these resources reach their maximum at $\varphi=\pi/2$. We demonstrate that classical correlations can significantly delay the decay of quantum correlations and coherence. This study underscores the importance of understanding the interplay between classical and quantum correlations in high-energy particle physics, particularly in the context of hyperon-antihyperon interactions explored in the BESIII experiment. This result could have potential applications in quantum information processing and high-energy physics.

\end{abstract}
\maketitle
\section{Introduction}    \label{sec:1}

Einstein, Podolsky, and Rosen (EPR) famously challenged the completeness of quantum mechanics in their 1935 thought experiment. This paradox highlighted the presence of non-local correlations between quantum systems, suggesting that a measurement on one system could instantaneously affect the state of another system, even if they are spatially separated \cite{ref1}. In response, Schr\"odinger introduced the concepts of entanglement and quantum steering, which provided a profound explanation of the non-local correlations intrinsic to quantum mechanics \cite{ref2}. Bell's theorem further demonstrated that local hidden variable theories fail to account for all nonlocal correlations observed between spatially separated systems under local quantum measurements \cite{ref3}. Entanglement, a purely quantum phenomenon with no classical equivalent, establishes correlations between parts of a multipartite quantum system. This unique resource is essential for numerous tasks in quantum information science \cite{ref4}. Beyond entanglement, other quantum correlations, including quantum non-locality without entanglement, have been shown to be valuable for quantum technology \cite{ref5,ref6}. Theoretical \cite{ref7} and experimental \cite{ref8} studies have shown that some separable states can outperform classical counterparts in specific tasks. This necessitates a broader investigation of quantum correlations.

Until recent years, it was widely believed that quantum correlations were closely tied to quantum entanglement, and the manipulation of quantum information was generally addressed within the framework of entanglement and separability \cite{ref8}. However, various studies have shown that entanglement is not the only type of correlation useful for the implementation of quantum protocols, and some separable states can also outperform their classical counterparts \cite{ref9,ref10}. This has led to the development of new quantifiers to detect nonclassical correlations beyond entanglement, thereby enhancing our ability to manipulate quantum information \cite{ref11}. The steerability of the two entangled bipartite states is assessed using quantum steering. This quantifier highlights the asymmetric relationship between Alice and Bob, two entangled observers. By exploiting their shared entanglement, Alice can thus modify (i.e., "steer") Bob's state \cite{ref14,refA}. However, these nonclassical correlations are sensitive to environmental disturbances, leading to decoherence, which disrupts quantum states and facilitates the transition to classical behavior \cite{ref15}.

The Geometric Quantum Discord (GQD) is a measure of quantum correlations in bipartite systems, designed to capture more general non-classical correlations beyond quantum entanglement. Geometric methods are frequently employed to define and measure quantum resources in a multitude of quantum systems \citep{ref16}. Specifically, the Schatten 1-norm quantum discord \citep{ref17,ref18}, is a reliable geometric-based indicator used to assess the level of quantum correlations in metal complexes \citep{ref19}. Unlike other correlation measures, it takes into account the geometry of the quantum state space. This approach allows quantifying quantum correlations in a way that considers the geometric structure of the quantum state space, which can provide richer information about quantum interactions. GQD has been widely studied both theoretically and experimentally, offering interesting prospects for the characterization and utilization of quantum resources in various contexts of quantum information processing and quantum communication \citep{ref20,ref21}.

Quantum coherence is indeed a fundamental concept in quantum physics, enabling a system to exist in a superposition of states, which is essential for various quantum phenomena, including interference and quantum optics \citep{ref22,ref23}. This property is not only crucial for theoretical understanding but also constitutes a valuable resource in many areas of research. For example, in quantum thermodynamics \citep{ref24,ref25}, quantum dots \citep{ref26,ref27}, Heisenberg spin chains \citep{ref28,ref29}, spin waves \citep{ref30}, quantum batteries \citep{ref31}, and even for developing algorithms for quantum computing \citep{ref32,ref33}. These applications highlight the importance of quantum coherence in many areas of quantum physics and its technological applications.

Studies on quantum entanglement have been conducted with low-energy protons in \cite{ref34}, while \cite{ref35} explores the possibility of achieving entanglement at colliders through the analysis of hadronic final states. The investigation of quantum entanglement extends to even smaller length scales at higher energies, as demonstrated in \cite{ref36}. In the field of high-energy physics, various experimental approaches have been proposed to investigate quantum entanglement, such as those utilizing neutral kaon physics \cite{ref37,ref38,ref39,ref40}, positronium \cite{ref41}, flavor oscillations in neutral B mesons \cite{ref42}, charmonium decays \cite{ref43}, and neutrino oscillations \cite{ref44}. Recently, the entanglement in top quark pair
production is observation at the LHC \cite{ref45}. Furthermore, \cite{ref46} established the experimental feasibility of observing Bell inequality violations within this system. In this regard, the study of entanglement has been proposed in high energy physics, including top quark production \cite{ref47,ref48,ref49,ref50,ref51}, hyperon production \cite{ref52}, and the production of gauge bosons, whether through Higgs boson decay \cite{ref53,ref54,ref55,ref56} or direct production mechanisms \cite{ref55,ref56}.

A novel approach to studying the decays of strange baryons is based on the investigation of hyperon-antihyperon pairs produced from \(J/\psi\) resonances in electron-positron collisions \cite{ref57}. The angular distribution of these processes can be compactly described using real-valued matrices, which represent both the initial spin-entangled state of the baryon-antibaryon pair and the weak two-body decay chains. These matrices, being modular, can be rearranged to model various decay scenarios, such as the processes \(e^+e^- \to \Lambda\bar{\Lambda}\) and similar ones \cite{ref57,ref58,ref59,ref60}. The BESIII experiment, conducted with an electron-positron collider, has performed extensive analyses of these angular distributions using multidimensional maximum-likelihood fits within this modular framework \cite{ref61,ref62}. These studies have opened new avenues for exploring fundamental aspects of quantum mechanics, such as nonlocality and quantum entanglement \cite{ref63,ref64}. Furthermore, the interest in studying quantum correlations in hyperon-antihyperon systems has been growing \cite{ref65,ref66}, particularly thanks to recent technological advancements in the BESIII detector. These upgrades enable more precise observations of entangled phenomena arising from electron-positron collisions \cite{ref67,ref68}.

The charmonium states, particularly $J/\psi$ and $\psi(2S)$, are vital sources of spin-entangled hyperon-antihyperon pairs, which are essential for studying the hyperon decay parameters and $\mathcal{CP}$ symmetry violations in the baryon sector \cite{ref69}. The recent BESIII experiment has made significant progress by observing polarization in the decay process $e^+e^- \to J/\psi \to \Lambda\bar{\Lambda}$, which allowed for the simultaneous determination of the decay asymmetries for $\Lambda$ and $\bar{\Lambda}$ \cite{ref70, ref71}. Notably, the recently measured decay asymmetry parameter for $\Lambda \to p\pi^-$ reveals significant deviations from previous values, suggesting that prior assessments of the polarization of $\Lambda$ and $\bar{\Lambda}$ may have been overestimated by approximately ($17\pm 3)$\% \cite{ref58}. These results emphasize the need to reevaluate the angular distributions and decay asymmetries in related weak decay processes, thereby enhancing our understanding of strong and weak interactions in particle physics.

In this study, we examine the quantum coherence and quantum correlations in the process $e^{+}e^{-} \to J/\psi \to \Lambda\bar{\Lambda}$ during the BESIII experiment, where $\Lambda$ and $\bar{\Lambda}$ represent the spin-$1/2$ hyperon and its antihyperon, respectively. Our analysis is based on the two-qubit density operator for the $\Lambda\bar{\Lambda}$ system. We show that quantum coherence and quantum correlations depend on the scattering angle $\varphi$. Furthermore, we illustrate the decoherence dynamics of the hyperon-antihyperon system $\Lambda\bar{\Lambda}$ when subjected to a classically correlated dephasing channel. Specifically, describing a channel as classically correlated refers to the existence of classical correlations between its successive actions on the hyperon and the antihyperon $\Lambda\bar{\Lambda}$.  Within this framework, we demonstrate that these classical correlations effectively delay the decay of quantum correlations and coherence.

The article is organized as follows: in section \ref{sec:2}, we describe the physical model used for a $\Lambda\bar{\Lambda}$ pair. Section \ref{sec:3} provides an overview of the concepts related to quantum steering, entanglement of formation, geometric quantum discord, and quantum coherence, considering the $\Lambda\bar{\Lambda}$ system in both Markovian and non-Markovian regimes. A comparison of the quantum resources between these two regimes is presented in section \ref{sec:4}, finally, the conclusions are summarized in section \ref{sec:5}.

\section{Theoretical model}\label{sec:2}

\begin{figure*}[t]
\begin{center}
\includegraphics[width=8cm,height=5.5cm]{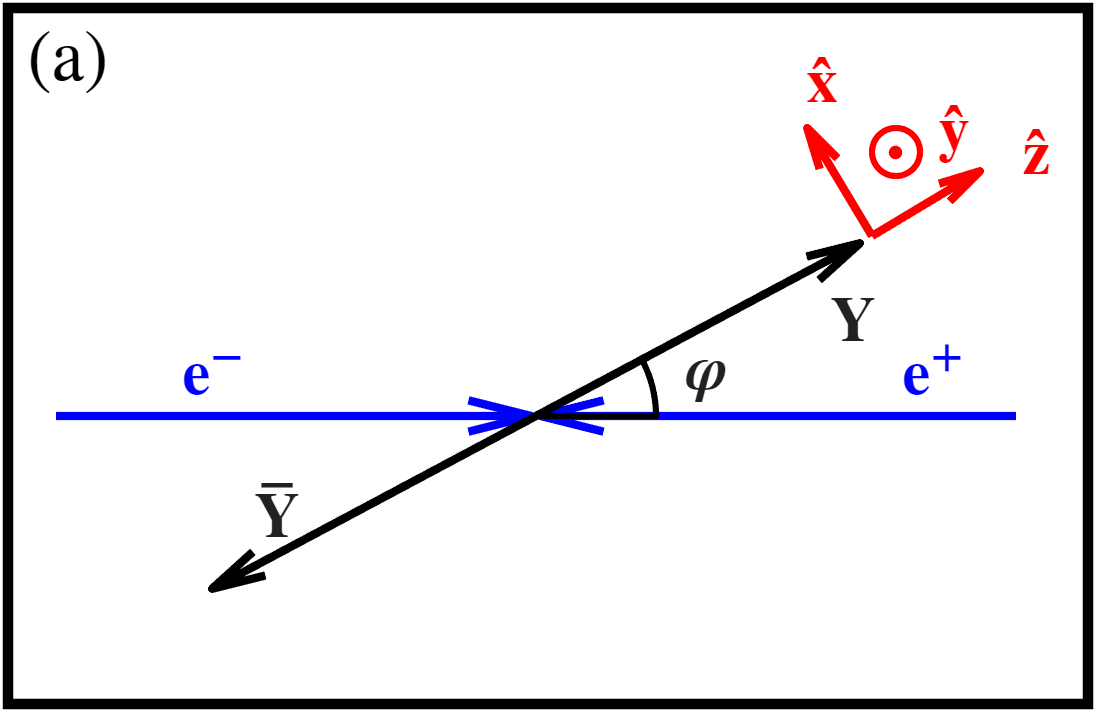}
\includegraphics[width=8cm,height=5.5cm]{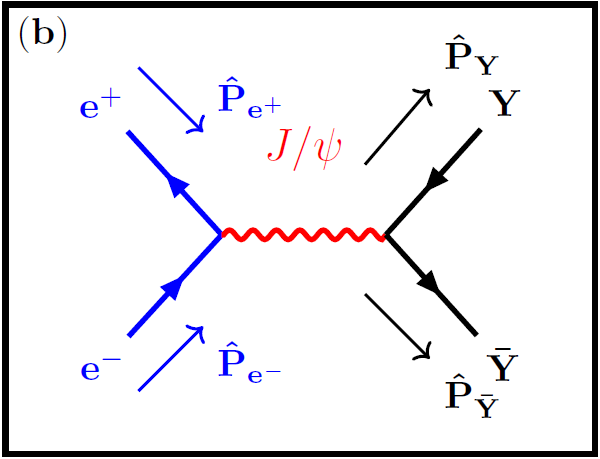}
\end{center}
\caption{(a) Coordinate system \(\{\boldsymbol{\hat{\rm x}}, \boldsymbol{\hat{\rm y}}, \boldsymbol{\hat{\rm z}}\}\) used in the rest frame of both \(\text{Y}\) and \(\bar{\text{Y}}\). (b) Feynman diagram representing the scattering process \(e^{+}e^{-} \to J/\psi \to \text{Y}\bar{\text{Y}}\).}
\label{fig:e}
\end{figure*}

Vector charmonia, including $J/\psi$ or $\psi(2S)$, are observed to decay into hyperon-antihyperon pairs, forming a massive two-qubit system $\text{Y} \bar{\text{Y}}$, consisting of two spin-$1/2$ particles. The hyperon and antihyperon are emitted with opposite momenta in the center-of-mass ($\mathcal{CP}$) frame due to momentum conservation. A two-qubit density operator describes the spin state of the hyperon-antihyperon pair \citep{ref58}
\begin{equation}
\begin{aligned}
\hat{\varrho}_{\text{Y}\bar{\text{Y}}} & := \frac{1}{4} \left[ \hat{\rm I} \otimes \hat{\rm I} + \sum_{i} \hat{\rm P}_i^{+} \left( \sigma_i \otimes \hat{\rm I} \right) + \sum_{j} \hat{\rm P}_j^{-} \left(\hat{\rm I} \otimes \sigma_j \right) + \sum_{i,j} \hat{\rm C}_{i,j} \left( \sigma_i \otimes \sigma_j \right) \right],
\end{aligned}
\label{eq:1}
\end{equation}
with $\boldsymbol{\sigma}  = (\sigma_x, \sigma_y, \sigma_z)$ denoting the Pauli matrices, $\hat{\rm P}^{\pm}$ representing the polarization (Bloch) vectors of the hyperon and antihyperon, respectively, and $\hat{\rm C}_{i,j}$ being their correlation matrix. The density matrix $\hat{\varrho}_{\text{Y}\overline{\text{Y}}}$ (Eq. (\ref{eq:1})), can be re-written as
\begin{equation}
\hat{\varrho}_{\text{Y}\bar{\text{Y}}} := \frac{1}{4}\sum_{\alpha,\beta=0}^{3}\Phi_{\alpha,\beta}\sigma^{\text{Y}}_{\alpha} \otimes \sigma^{\bar{\text{Y}}}_{\beta},
\end{equation}
where $\Phi_{0,0} = 1$, $\Phi_{i,0} = \hat{\rm P}_i^{+}$, $\Phi_{0,j} = \hat{\rm P}_j^{-}$, $\Phi_{i,j} = \hat{\rm P}_{i,j}$, and $\sigma_0={\rm diag}(1,1)$. A set of four Pauli matrices $\sigma^{\text{Y}}_{\alpha}$($\sigma^{\bar{\text{Y}}}_{\beta}$) acting in the rest frame of a hyperon $\text{Y}$($\bar{\text{Y}}$) is used, and $\Phi_{\alpha,\beta}$ is a $4 \times 4$ real matrix representing the polarization and spin correlations of the hyperon. For the hyperon $\text{Y}$, we define its helicity rest frame as (Fig. \ref{fig:e}(a))
\begin{equation}
\hat{\boldsymbol{\rm y}} = \frac{\hat{\boldsymbol{\rm P}}_e \times \hat{\boldsymbol{\rm P}}_\text{Y}}{|\hat{\boldsymbol{\rm P}}_e \times \hat{\boldsymbol{\rm P}}_\text{Y} |}, \quad \hat{\boldsymbol{\rm z}} = \boldsymbol{\hat{\rm P}}_\text{Y}, \quad \boldsymbol{\hat{\rm x}} = \boldsymbol{\hat{\rm y}} \times \boldsymbol{\hat{\rm z}}.
\label{eq:haty}
\end{equation}

For the antihyperon $\bar{\text{Y}}$, we also adopt its rest frame, with the axes chosen to be the same as those of the hyperon's: $\left\lbrace \boldsymbol{\hat{\rm x}}_{\bar{\text{Y}}}, \boldsymbol{\hat{\rm y}}_{\bar{\text{Y}}}, \boldsymbol{\hat{\rm z}}_{\bar{\text{Y}}} \right\rbrace =\left\lbrace \boldsymbol{\hat{\rm x}},\boldsymbol{\hat{\rm y}},\boldsymbol{\hat{\rm z}}\right\rbrace$. The three axes we have chosen are the same as those in Ref. \cite{ref64}, which is shown in Fig. \ref{fig:e}(a).
The elements of the matrix $\hat{\rm C}_{i,j}$ are functions of the production angle $\varphi$, which is the angle between the momenta of the incoming electron and the outgoing hyperon, and $\cos(\varphi) = \boldsymbol{\hat{\rm P}}_e \cdot \boldsymbol{\hat{\rm P}}_\text{Y}$. In the rest frames of $\text{Y}$ and $\bar{\text{Y}}$, The form of $\Phi_{\alpha,\beta}$ is determined by virtual photon exchange \cite{ref58}, using the coordinate system defined in Eq. (\ref{eq:haty}) 
\begin{equation}
\Phi_{\alpha,\beta}:= 
\begin{bmatrix}
1  & 0 & \hat{\rm P}^{+}_{\rm y} & 0 \\
0 & \hat{\rm C}_{\rm x,\rm x} & 0 & \hat{\rm C}_{\rm x,\rm z} \\
\hat{\rm P}^{-}_{\rm y} & 0 & \hat{\rm C}_{\rm y,\rm y} & 0 \\
0 & \hat{\rm C}_{\rm z,\rm x} & 0 & \hat{\rm C}_{\rm z,\rm z}
\end{bmatrix}.
\label{eq:3}
\end{equation}

The polarization and correlations can be extracted from $\Phi_{\alpha,\beta}$ in Eq. (\ref{eq:3}) as follows
\begin{equation*}
\begin{aligned}
\hat{\rm P}_{\rm y}^{+} &= \hat{\rm P}_{\rm y}^{-} = \frac{\sqrt{1 - \upsilon_\psi^2} \sin(\Delta\theta)\sin(\varphi)\cos(\varphi)}{1+\upsilon_{\psi}\cos^2(\varphi)}, \\  
\hat{\rm C}_{\rm x,\rm x} &= \frac{\sin^2(\varphi)}{1+\upsilon_{\psi}\cos^2(\varphi)}, \quad
\hat{\rm C}_{\rm y,\rm y} = \frac{-\upsilon_{\psi}\sin^2(\varphi)}{1+\upsilon_{\psi}\cos^2(\varphi)}, \quad
\hat{\rm C}_{\rm z,\rm z} = \frac{\upsilon_{\psi} + \cos^2(\varphi)}{1+\upsilon_{\psi}\cos^2(\varphi)}, \\  
\hat{\rm C}_{\rm x,\rm z} &= \hat{\rm C}_{\rm z,\rm x} = \frac{\sqrt{1 - \upsilon_\psi^2} \cos(\Delta\theta)\sin(\varphi)\cos(\varphi)}{1+\upsilon_{\psi}\cos^2(\varphi)},
\end{aligned}
\end{equation*}
where $\hat{\rm P}_{\rm y}^{+}$ and $\hat{\rm P}_{\rm y}^{-}$ denote the polarizations of the $\text{Y}$ and $\bar{\text{Y}}$ particles along the $\hat{\boldsymbol{\rm y}}$ direction, which is normal to the production plane. $\upsilon_\psi \in [-1,1]$ is the decay parameter of the vector charmonium $\psi(c\bar{c})$ and $\Delta\theta \in [-\pi,\pi]$ is the relative form factor phase. The symmetry properties of the polarization and correlation are derived from parity and charge conjugation invariance, with the assumption of no $\mathcal{CP}$ violation.\\
The transformation of the two-qubit state from Eqs. (\ref{eq:1}) and (\ref{eq:3}) into an X-state (represents the symmetric X-state form for $\hat{\varrho}_{\text{Y}\bar{\text{Y}}}$), achieved by swapping the $\hat{\boldsymbol{\rm y}}$ and $\hat{\boldsymbol{\rm z}}$ axes and diagonalizing $\hat{\rm C}_{\rm i\rm j}$, simplifies the analysis, as discussed in \cite{X1}. The resulting spin density matrix is
\begin{equation} \label{varrhoX}
\hat{\varrho}_{\text{Y}\overline{\text{Y}}}^X = \frac{1}{4} \left(\hat{\rm I} \otimes \hat{\rm I} + \kappa \sigma_z \otimes \hat{\rm I} + \hat{\rm I} \otimes \kappa \sigma_z + \sum_{i=1}^{3} \gamma_i \sigma_i \otimes \sigma_i \right).
\end{equation}

The matrix $\Phi_{\alpha\beta}$ for this state becomes
\begin{equation}
\Phi^X_{\alpha,\beta} =
\begin{bmatrix}
1 & 0 & 0 & \kappa \\
0 & \gamma_1 & 0 & 0 \\
0 & 0 & \gamma_2 & 0 \\
\kappa & 0 & 0 & \gamma_3
\end{bmatrix},
\label{eq:Theta}
\end{equation}
where
\begin{equation*}
\begin{aligned}
\kappa &= \frac{\sqrt{1 - \upsilon_\psi^2} \sin(\Delta\theta)\sin(\varphi)\cos(\varphi)}{1+\upsilon_{\psi} \cos^2(\varphi)}, \\  
\gamma_{1,2} &= \frac{1 + \upsilon_{\psi} \pm \sqrt{\bigg(1 +\upsilon_{\psi} \cos^2(2\varphi)\bigg)^2 - \big(1 -\upsilon_\psi^2\big) \sin^2(\Delta\theta) \sin^2(2\varphi)}}{2(1+\upsilon_{\psi}\cos^2(\varphi))}, \\  
\gamma_3 &= \frac{-\upsilon_{\psi}\sin^2(\varphi)}{1 + \upsilon_{\psi} \cos^2(\varphi)}.
\end{aligned}
\end{equation*}

Using Eq. (\ref{varrhoX}), the spin density operator for the hyperon-antihyperon system can be rewritten in the $\sigma_z$ basis
\begin{equation}
\tilde{\varrho}_{\text{Y}\bar{\text{Y}}}^{X} =
\begin{bmatrix}
\hat{\varrho}_{1,1}& 0 & 0 & \hat{\varrho}_{1,4} \\
0 & \hat{\varrho}_{2,2} & \hat{\varrho}_{2,3} & 0 \\
0 & \hat{\varrho}_{2,3} & \hat{\varrho}_{2,2}& 0 \\
\hat{\varrho}_{1,4} & 0 & 0 & \hat{\varrho}_{4,4}
\end{bmatrix},
\label{eq:rho}
\end{equation}
where

\begin{equation*}
\begin{aligned}
\hat{\varrho}_{1,1}& = \frac{1}{4} \Bigg(1 + \frac{2\sqrt{1 -\upsilon_\psi^2} \sin(\Delta\theta)\sin(\varphi)\cos(\varphi)}{1+\upsilon_{\psi}\cos^{2}(\varphi)} -\frac{\upsilon_{\psi}\sin^2(\varphi)}{1 + \upsilon_{\psi}\cos^2(\varphi)}\Bigg), \\
\hat{\varrho}_{1,4}&=\frac{\sqrt{\bigg(1 +\upsilon_{\psi} \cos^2(2\varphi)\bigg)^2 - \big(1 - \upsilon_\psi^2\big)\sin^2(\Delta\theta) \sin^2(2\varphi)}}{4\bigg(1+\upsilon_{\psi}\cos^2(\varphi)\bigg)}, \\
\hat{\varrho}_{2,2}&= \hat{\varrho}_{2,3}= \frac{1+\upsilon_{\psi}}{4\bigg(1 +\upsilon_{\psi}\cos^2(\varphi)\bigg)}, \\
\hat{\varrho}_{4,4}&= \frac{1}{4} \Bigg(1 - \frac{2\sqrt{1 -\upsilon_\psi^2} \sin(\Delta\theta)\sin(\varphi)\cos(\varphi)}{1+\upsilon_{\psi}\cos^{2}(\varphi)} -\frac{\upsilon_{\psi}\sin^2(\varphi)}{1 + \upsilon_{\psi}\cos^2(\varphi)}\Bigg).
\end{aligned}
\end{equation*}

\section{Preliminary of correlated quantum channel}\label{sec:3}

This section summarizes the preliminary concepts relevant to characterizing correlated quantum channels. To simplify the analysis, we consider a two-qubit system that is initially prepared in the state $\hat{\varrho}(0)$. We consider two qubits passing through channel $\zeta$, resulting in an output state described by a map if the channel acts identically on each qubit \cite{p4}
\begin{equation}
\hat{\varrho}(t)= \zeta[\hat{\varrho}(0)]=\sum_{i,j=0}^{3}{\rm L}_{i,j}\hat{\varrho}(0){\rm L}_{i,j}^{\dagger},
\label{eq:D}
\end{equation}
with Kraus operators ${\rm L}_{i,j}$ writes as
\begin{equation}
{\rm L}_{i,j}=\sqrt{{\rm p}_{i,j}} \sigma_{i}\otimes\sigma_{j},
\end{equation}
where $\sigma_{0}={\rm diag}(1,1)$ is the $2\times 2$ identity matrix, and $\sigma_{x,y,z}$ denote the Pauli operators. The values ${\rm p}_{i}$ form a probability distribution, meaning each ${\rm p}_{i}$ is non-negative and their sum is 1. The operators ${\rm L}_{i,j}$ are completely positive and trace-preserving (CPTP), satisfying $\sum_{i,j}{\rm L}_{i,j}=\hat{\rm I}_{4}$. The joint probability is given by
\begin{equation}
{\rm p}_{i,j}=(1-\mu){\rm p}_{i}{\rm p}_{j}+\mu {\rm p}_{i}\delta_{i,j},
\label{eq:p}
\end{equation}
where $\delta_{i,j}$ is the Kronecker delta, and $0 \leq \mu \leq 1$ quantifies the classical correlations between consecutive applications of the operation $\zeta$ to the two qubits. This model incorporates the effects of standard uncorrelated dephasing. We consider the dephasing channel with probabilities ${\rm p}_{0}=1-{\rm p}$, ${\rm p}_{x,y}=0$, and ${\rm p}_{z}={\rm p}$. To explore the time-dependent quantum coherence, we examine a colored pure dephasing model with a time-dependent Hamiltonian \cite{p4}
\begin{equation}
\mathcal{H}(t)=\hbar\Gamma(t)\sigma_{z},
\end{equation}
where $\Gamma(t)=\omega {\rm m}(t)$ is a random telegraph signal, with m(t) following a Poisson distribution with mean $\left\langle {\rm m}(t)\right\rangle =t/(2\tau)$, and $\omega$ is a coin-flip random variable that can take either the value $\pm \omega$. For $\omega=1$, the time-dependent factor ${\rm p}$ is given by 
\begin{equation}
{\rm p}=\left[1-K(t)\right]/2
\end{equation}
In the non-Markovian regime, which is characterized by the condition $4\tau >1$, we have
\begin{equation}
K(t)=\e^{-ut}\left[\cos(vt)+\frac{1}{v}\sinh(vt) \right]
\label{eq:rtt}
\end{equation}
In the Markovian regime, characterized by $4\tau <1$, with $u=\frac{1}{2\tau}$ and $v=\sqrt{|u^{2}-1|}$, we have
\begin{equation}
K(t)=\e^{-ut}\left[\cosh(vt)+\frac{1}{v}\sin(vt) \right]
\label{eq:rt}
\end{equation}
The two-qubit density operator for the $\Lambda\bar{\Lambda}$ system after correlated dephasing is derived as follows. Starting with (\ref{eq:rho}) and applying (\ref{eq:D}) and (\ref{eq:p}), the output state is
\begin{equation}
\begin{aligned}
\hat{\varrho}_{\text{Y}\bar{\text{Y}}}(t)=
\begin{bmatrix}
\hat{\varrho}_{1,1}& 0 & 0 & \eta\hat{\varrho}_{1,4} \\
0 & \hat{\varrho}_{2,2} &\eta\hat{\varrho}_{2,3} & 0 \\
0 & \eta\hat{\varrho}_{3,2} & \hat{\varrho}_{3,3}& 0 \\
\eta\hat{\varrho}_{4,1} & 0 & 0 & \hat{\varrho}_{4,4}
\end{bmatrix},
\label{eq:varrho}
\end{aligned}
\end{equation}
where 
\begin{equation}
\eta = K^{2}(t)+\left[ 1-K^{2}(t)\right]\mu.
\label{eq:kappa}
\end{equation}

\section{Quantum correlation measures}\label{sec:3}

This section explores the use of the hyperon-antihyperon spin density operator to examine various quantum properties in the $\Lambda\bar{\Lambda}$ system, including quantum steering, entanglement of formation, geometric quantum discord, and quantum coherence. However, we have chosen to focus on $\Lambda$ because the results obtained from this decay are representative of the trends observed for other hyperons. In particular, the quantum correlations and coherence effects extracted from the decay of $\Lambda$ exhibit similarities with those arising from the decays of $\Xi^{-}$, $\Xi^{0}$, and $\Sigma$. This choice allows us to simplify the analysis without any loss of generality while highlighting the main physical effects studied in this work.
\begin{table}[H]
\begin{center}
\caption{Some parameters in $e^{+}e^{-} \rightarrow J/\psi \rightarrow \text{Y} \overline{\text{Y}}$, where $\text{Y} \overline{\text{Y}}$ is a pair of ground-state octet hyperons.}
\begin{tabular}{c c c c} 
\hline
\hline
& $\upsilon_{\psi}$ & $\Delta\theta/rad$ &  Ref  \\
\hline
 $J/\psi\rightarrow \Lambda\overline{\Lambda}$  & 0.475(4) & 0.752(8)& \cite{ref67,Ta2}\\
\hline
$J/\psi\rightarrow \Sigma^{+}\overline{\Sigma}^{-}$ &-0.508(7)  & -0.270(15) & \cite{Ta3,Ta4}\\
\hline
$J/\psi\rightarrow \Xi^{-}\overline{\Xi}^{+}$& 0.586(16)&1.213(49)& \cite{ref62,Ta6}\\
\hline
$J/\psi\rightarrow \Xi^{0}\overline{\Xi}^{0}$ &0.514(16)&1.168(26)& \cite{Ta7,Ta8} \\
\hline
\end{tabular}
\end{center}
\end{table}
This similarity has been confirmed through comparative analyses and is consistent with previous experimental/theoretical results \cite{ref58}.

\subsection{Quantum steering}

Quantum steering describes correlations where measurements on one particle (at $\Lambda$) seem to instantaneously influence the state of a distant particle (at $\bar{\Lambda}$). Given a shared two-qubit state $\hat{\varrho}_{\Lambda\bar{\Lambda}}$ between Alice and Bob, we examine the possibility of steering. Alice can demonstrate steering to Bob if the steering operators $\hat{\tau}_{\Lambda \bar{\Lambda}}$ and $\hat{\tau}_{\bar{\Lambda} \Lambda}$ are defined as follows \cite{s6}
\begin{equation}
\hat{\tau}_{\Lambda\bar{\Lambda}}=\frac{1}{\sqrt{3}}\hat{\varrho}_{\Lambda\bar{\Lambda}}+\left(1-\frac{1}{\sqrt{3}}\right)\hat{\varrho}_{\bar{\Lambda}}
\label{eq:s}
\end{equation}
and 
\begin{equation}
\hat{\tau}_{\bar{\Lambda}\Lambda}=\frac{1}{\sqrt{3}}\hat{\varrho}_{\Lambda\bar{\Lambda}}+\left(1-\frac{1}{\sqrt{3}}\right)\hat{\varrho}_{\Lambda},
\end{equation}

where $\hat{\varrho}_{\bar{\Lambda}}=\frac{\mathbb{I}}{2}\otimes \varrho_{\bar{\Lambda}}$ and $\hat{\varrho}_{\Lambda}=\varrho_{\Lambda}\otimes\frac{\mathbb{I}}{2}$, with $\varrho_{\bar{\Lambda}}=\textmd{tr}_{\Lambda}(\hat{\varrho}_{\Lambda\bar{\Lambda}})$ and $\varrho_{\Lambda}=\textmd{tr}_{\bar{\Lambda}}(\hat{\varrho}_{\Lambda\bar{\Lambda}})$ being the reduced states for Bob and Alice, respectively. By simple calculation, we find that the matrix $\hat{\tau}_{\bar{\Lambda}\Lambda}$ for an X-state is
{\small 
\begin{equation}
\hat{\tau}_{\bar{\Lambda}\Lambda}=
\begin{pmatrix}
\frac{\sqrt{3}}{3}\hat{\varrho}_{1,1}+s& 0 & 0 & \frac{\sqrt{3}}{3}\eta\hat{\varrho}_{1,4} \\
0 &\frac{\sqrt{3}}{3}\hat{\varrho}_{2,2}+s& \frac{\sqrt{3}}{3}\eta\hat{\varrho}_{2,3}& 0 \\
0 & \frac{\sqrt{3}}{3}\eta\hat{\varrho}_{2,3} & \frac{\sqrt{3}}{3}\hat{\varrho}_{3,3}+q& 0 \\
\frac{\sqrt{3}}{3}\eta\hat{\varrho}_{1,4} & 0 & 0 & \frac{\sqrt{3}}{3}\hat{\varrho}_{4,4}+q
\end{pmatrix},
\end{equation}}
where $s=\frac{3-\sqrt{3}}{6}(\hat{\varrho}_{1,1}+\hat{\varrho}_{2,2})$ and $q=\frac{3-\sqrt{3}}{6}(\hat{\varrho}_{3,3}+\hat{\varrho}_{4,4})$. 
 
The above state, $\hat{\tau}_{\bar{\Lambda}\Lambda}$ is entangled if it satisfies one of the following inequalities \cite{s6,s4} 
\begin{subequations}
\begin{align}
|\eta\hat{\varrho}_{1,4}|^{2}>f_{a}-f_{b},
\end{align}
\begin{align}
|\eta\hat{\varrho}_{2,3}|^{2}>f_{c}-f_{b},
\end{align}
\end{subequations}
where
\begin{subequations}
\begin{align}
f_{a}=&\frac{2-\sqrt{3}}{2}\hat{\varrho}_{1,1}\hat{\varrho}_{4,4}+\frac{2+\sqrt{3}}{2}\hat{\varrho}_{2,2}\hat{\varrho}_{3,3}+\frac{1}{4}(\hat{\varrho}_{1,1}+\hat{\varrho}_{4,4})(\hat{\varrho}_{2,2}+\hat{\varrho}_{3,3}),
\label{eq:a}
\end{align}
\begin{align}
f_{b}=\frac{1}{4}(\hat{\varrho}_{1,1}-\hat{\varrho}_{4,4})(\hat{\varrho}_{2,2}-\hat{\varrho}_{3,3}),
\label{eq:b}
\end{align}
\begin{align}
f_{c}=&\frac{2+\sqrt{3}}{2}\hat{\varrho}_{1,1}\hat{\varrho}_{4,4}+\frac{2-\sqrt{3}}{2}\hat{\varrho}_{2,2}\hat{\varrho}_{3,3}+\frac{1}{4}(\hat{\varrho}_{1,1}+\hat{\varrho}_{4,4})(\hat{\varrho}_{2,2}+\hat{\varrho}_{3,3}).
\label{eq:c}
\end{align}
\end{subequations}
Similarly, steering from Alice to Bob is verified by one of the following inequalities
\begin{subequations}
\begin{align}
|\eta\hat{\varrho}_{1,4}|^{2}>f_{a}+f_{b},
\label{eq:I1}
\end{align}
\begin{align}
|\eta\hat{\varrho}_{2,3}|^{2}>f_{c}+f_{b}.
\label{eq:I2}
\end{align}
\end{subequations}
The steerability of Bob to Alice, denoted by $\mathcal{S}_{\bar{\Lambda}\Lambda}$, is given by  \cite{s6}
{\small \begin{equation}\label{AtoB}
\mathcal{S}_{\bar{\Lambda}\Lambda}=\max\left\lbrace 0,~\frac{8}{\sqrt{3}}\left[|\eta\hat{\varrho}_{1,4}|^{2}-f_{a}+f_{b},~|\eta\hat{\varrho}_{2,3}|^{2}-f_{c}+f_{b}\right]\right\rbrace.
\end{equation}}

The steerability from Alice to Bob, $\mathcal{S}_{\Lambda\bar{\Lambda}}$, given by
{\small \begin{equation}\label{BtoA}
 \mathcal{S}_{\Lambda\bar{\Lambda}}=\max\left\lbrace 0,~\frac{8}{\sqrt{3}}\left[|\eta\hat{\varrho}_{1,4}|^{2}-f_{a}-f_{b},~|\eta\hat{\varrho}_{2,3}|^{2}-f_{c}-f_{b}\right]\right\rbrace. 
\end{equation}}
The steering asymmetry is defined as
\begin{equation}
\delta \mathcal{S}=\vert \mathcal{S}_{\Lambda\bar{\Lambda}}-\mathcal{S}_{\bar{\Lambda}\Lambda} \vert.
\end{equation}
The steerability between qubit ($\Lambda$) (Alice) and qubit ($\bar{\Lambda}$) (Bob) can be classified into distinct cases:
\begin{itemize}
    \item With $\delta \mathcal{S}> 0$ (one-way steering), either Alice can steer Bob ($\mathcal{S}_{\Lambda\bar{\Lambda}} > 0$ and $\mathcal{S}_{\bar{\Lambda}\Lambda} = 0$) or Bob can steer Alice ($\mathcal{S}_{\Lambda\bar{\Lambda}} = 0$ and $\mathcal{S}_{\bar{\Lambda}\Lambda} > 0$).   
    \item When $\delta \mathcal{S}= 0$ (two-way or no-way steering), either $\mathcal{S}_{\Lambda\bar{\Lambda}} = \mathcal{S}_{\bar{\Lambda}\Lambda} = 0$ (no steering) or $\mathcal{S}_{\Lambda\bar{\Lambda}} = \mathcal{S}_{\bar{\Lambda}\Lambda} > 0$ (two-way steering).   
\end{itemize}
We first examine the quantum steerability of the hyperon-antihyperon system, characterized by its spin density operator (Eq. \ref{eq:varrho}) as a function of the classical parameter $\mu$ (with $\varphi= \pi/2$) and the scattering angle $\varphi$ (with $\mu=0.8$). The condition $\hat{\varrho}_{2,2} = \hat{\varrho}_{3,3}$ leads to $f_{b}=0$ according to Eq. (\ref{eq:b}). As a result, the steerabilities from $\bar{\Lambda}$ to $\Lambda$ (Eq. (\ref{AtoB})) and from $\Lambda$ to $\bar{\Lambda}$ (Eq. (\ref{BtoA})) are equal, i.e., $\mathcal{S}_{\Lambda\bar{\Lambda}}= \mathcal{S}_{\bar{\Lambda}\Lambda} = \max\left\lbrace 0,~\frac{8}{\sqrt{3}}\left[|\eta\hat{\varrho}_{1,4}|^{2} - f_{a},~|\eta\hat{\varrho}_{2,3}|^{2} - f_{c}\right]\right\rbrace.$ This equality implies that $\delta \mathcal{S}=0$, indicating that the steerability between $\Lambda$ and $\bar{\Lambda}$ is bidirectional (two-way or no-way steering).
\begin{figure}[!h]
\includegraphics[scale=0.45]{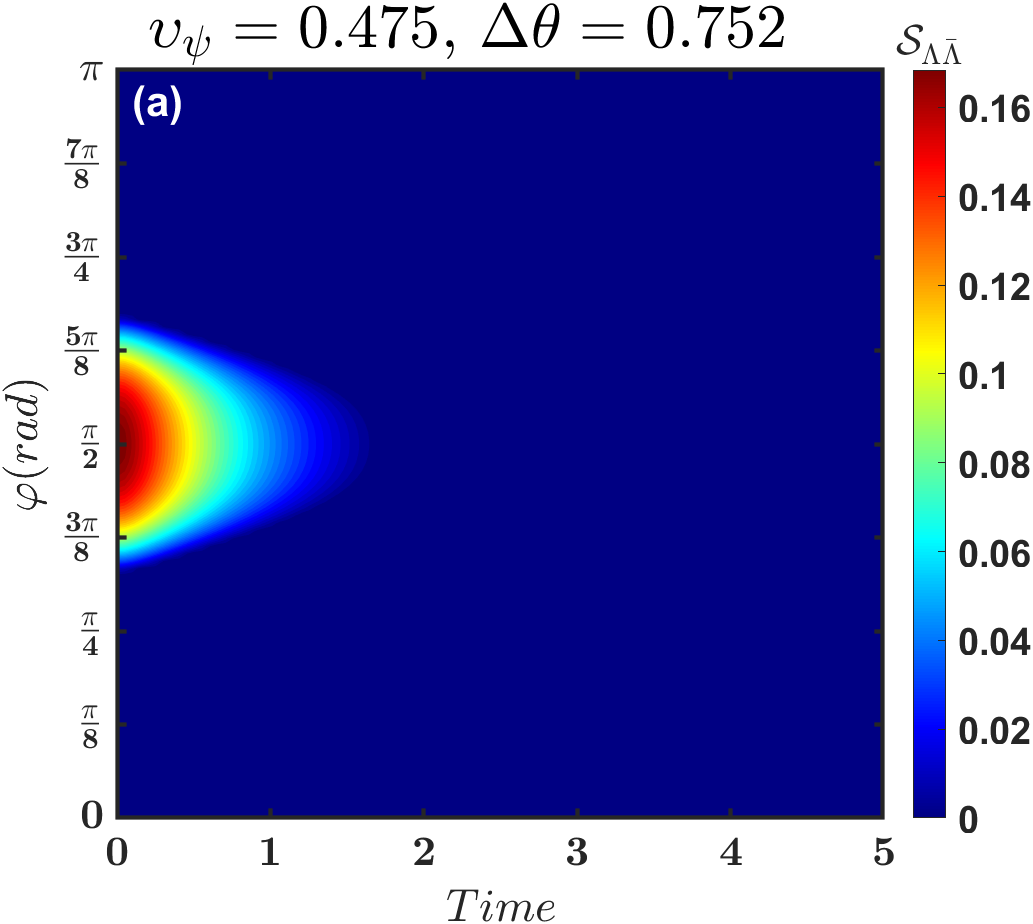}
\includegraphics[scale=0.45]{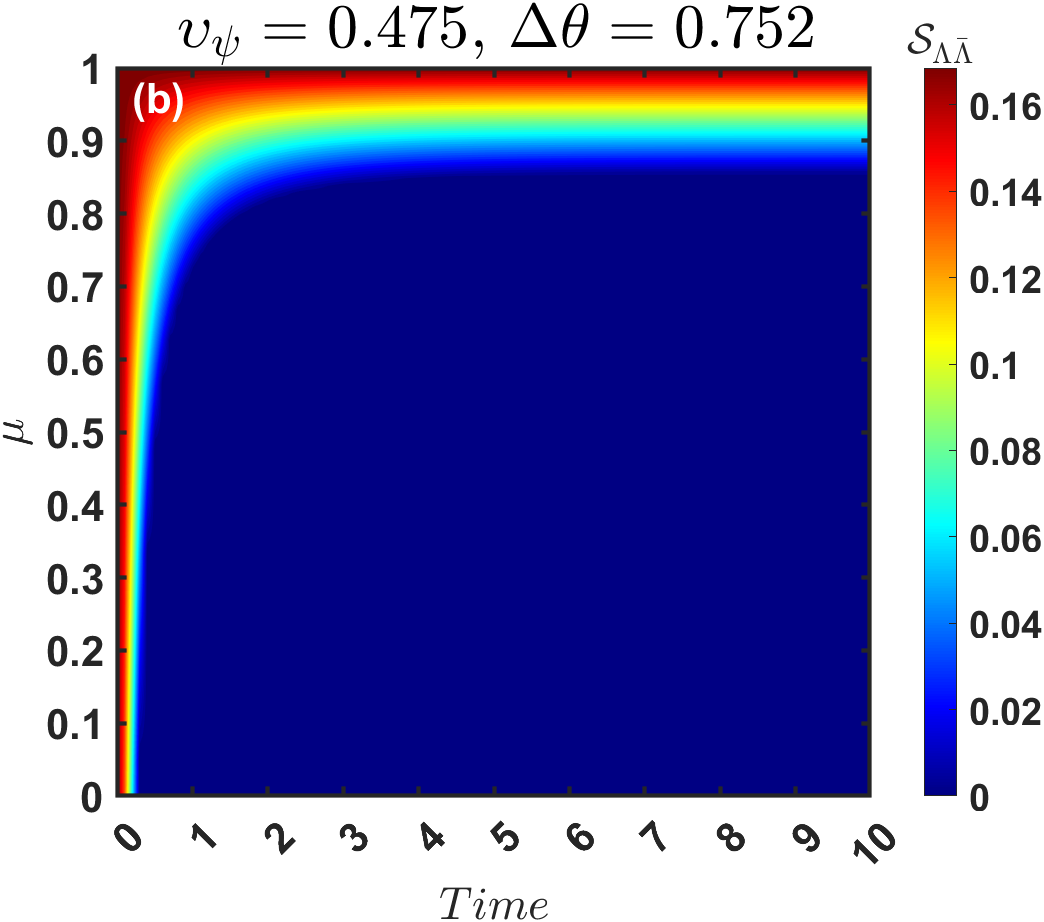}
\caption{Time evolution of quantum steering $\mathcal{S}_{\Lambda\bar{\Lambda}}$ in the Markovian regime ($\tau = 0.1$) is plotted for (a) $\mu = 0.8$ and (b) $\varphi=\pi/2$.}
\label{fig:h1}
\end{figure}

\begin{figure}[!h]
\includegraphics[scale=0.45]{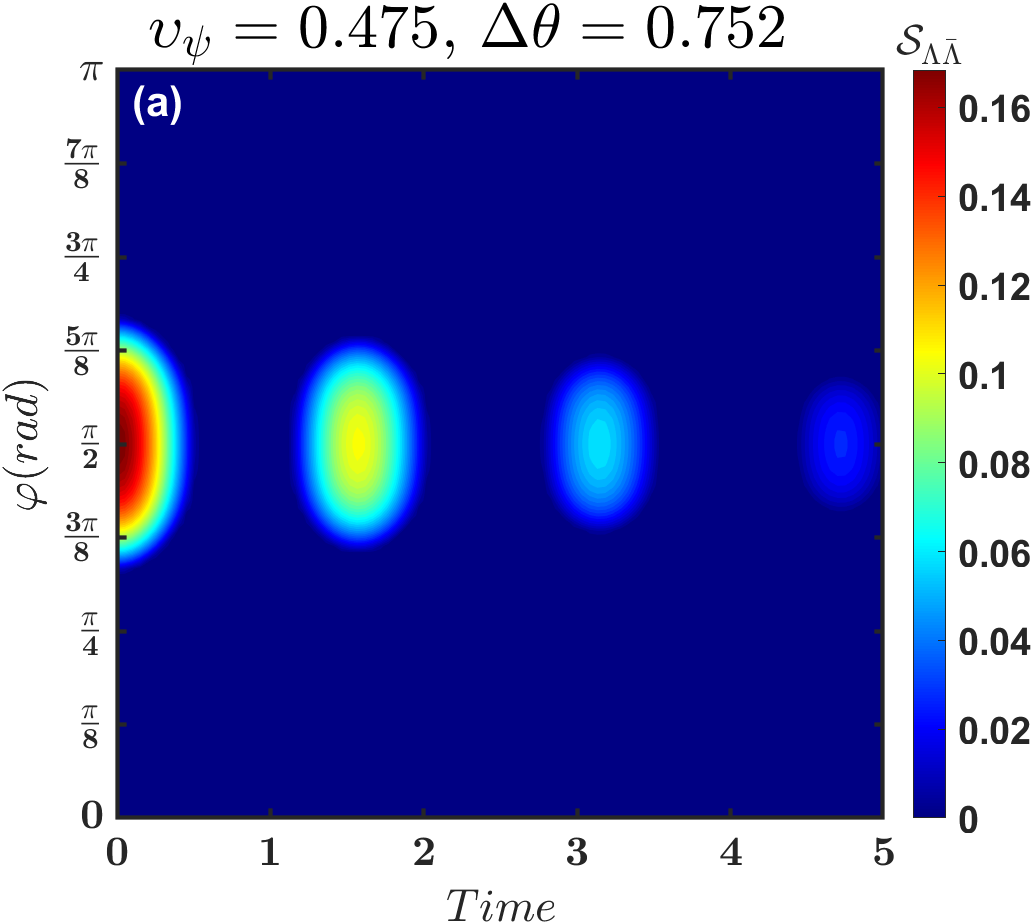}
\includegraphics[scale=0.45]{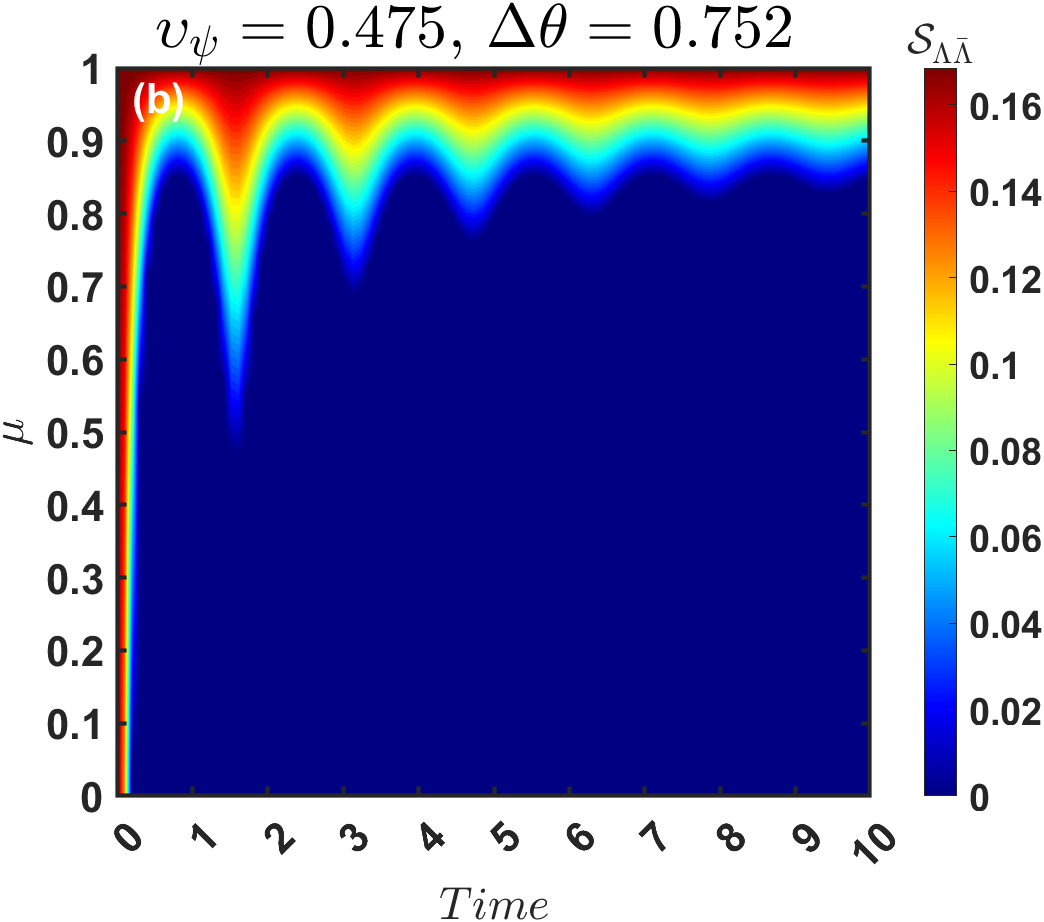}
\caption{Time evolution of quantum steering $\mathcal{S}_{\Lambda\bar{\Lambda}}$ in the non-Markovian regime ($\tau = 5$) is plotted for (a) $\mu = 0.8$ and (b) $\varphi=\pi/2$.}
\label{fig:h2}
\end{figure}  
Figure \ref{fig:h1}(a) shows the steerability $\mathcal{S}_{\Lambda\bar{\Lambda}}$ as a function of both time and the diffusion angle $\varphi$ in the Markovian regime. The steerability peaks at $\varphi=\pi/2$ and diminishes monotonically as time progresses or $\varphi$ moves away from $\pi/2$.

For $\varphi = 0$ and $\varphi= \pi$, the density matrix \(\hat{\varrho}_{\Lambda\bar{\Lambda}}\) is given by 
\begin{equation*}
\hat{\varrho}_{\Lambda\bar{\Lambda}}(t) =
\begin{bmatrix}
1 & 0 & 0 & \eta \\
0 & 1 & \eta & 0 \\
0 & \eta & 1 & 0 \\
\eta & 0 & 0 & 1
\end{bmatrix}.
\end{equation*} 
This form allows us to determine the values of \(f_{a}\), \(f_{b}\), and \(f_{c}\) in equations \eqref{eq:a}, \eqref{eq:b}, and \eqref{eq:c}, respectively
\[
f_{a} = f_{c} = 3,
\]
\[
f_{b} = 0.
\]
We observe that the inequalities given in equations \eqref{eq:I1} and \eqref{eq:I2} are not satisfied.\\
Moreover, the sterability between the qubits \(\Lambda\) and \(\bar{\Lambda}\) is found to be zero
\[
\mathcal{S}_{\Lambda\bar{\Lambda}} = \mathcal{S}_{\bar{\Lambda}\Lambda} = 0.
\]

Figure \ref{fig:h1}(b) explores the behavior of the steerability $\mathcal{S}_{\Lambda\bar{\Lambda}}$ as a function of time and the classical correlation parameter $\mu$ in the Markovian regime. The figure demonstrates that increasing $\mu$ leads to a significant enhancement of quantum steerability. Furthermore, the steerability grows over time, eventually reaching a steady state.

In Fig. \ref{fig:h2}(a), we plot the quantum steerability versus the time and scattering angle $\varphi$ in the non-Markovian regime. In contrast to the Markovian case, where steerability generally decreases monotonically with time, the non-Markovian regime exhibits periodic oscillations in steerability. Besides, quantum steerability reaches its maximum at $\varphi = \pi/2$. Moreover, the memory effects inherent to the non-Markovian regime induce a non-monotonic behavior of quantum steering over time.

We plot in Fig. \ref{fig:h2}(b) the quantum steerability vs time and the classical correlation parameter $\mu$ in a non-Markovian regime. We note that the amplitude of the quantum steerability oscillations decreases as the classical correlation parameter $\mu$ increases. Furthermore, the decay rate of steerability becomes significantly slower as $\mu$ takes higher values, supporting the idea that stronger classical correlations contribute to a better preservation of quantum steerability.
\begin{figure}[!h]
\includegraphics[scale=0.4]{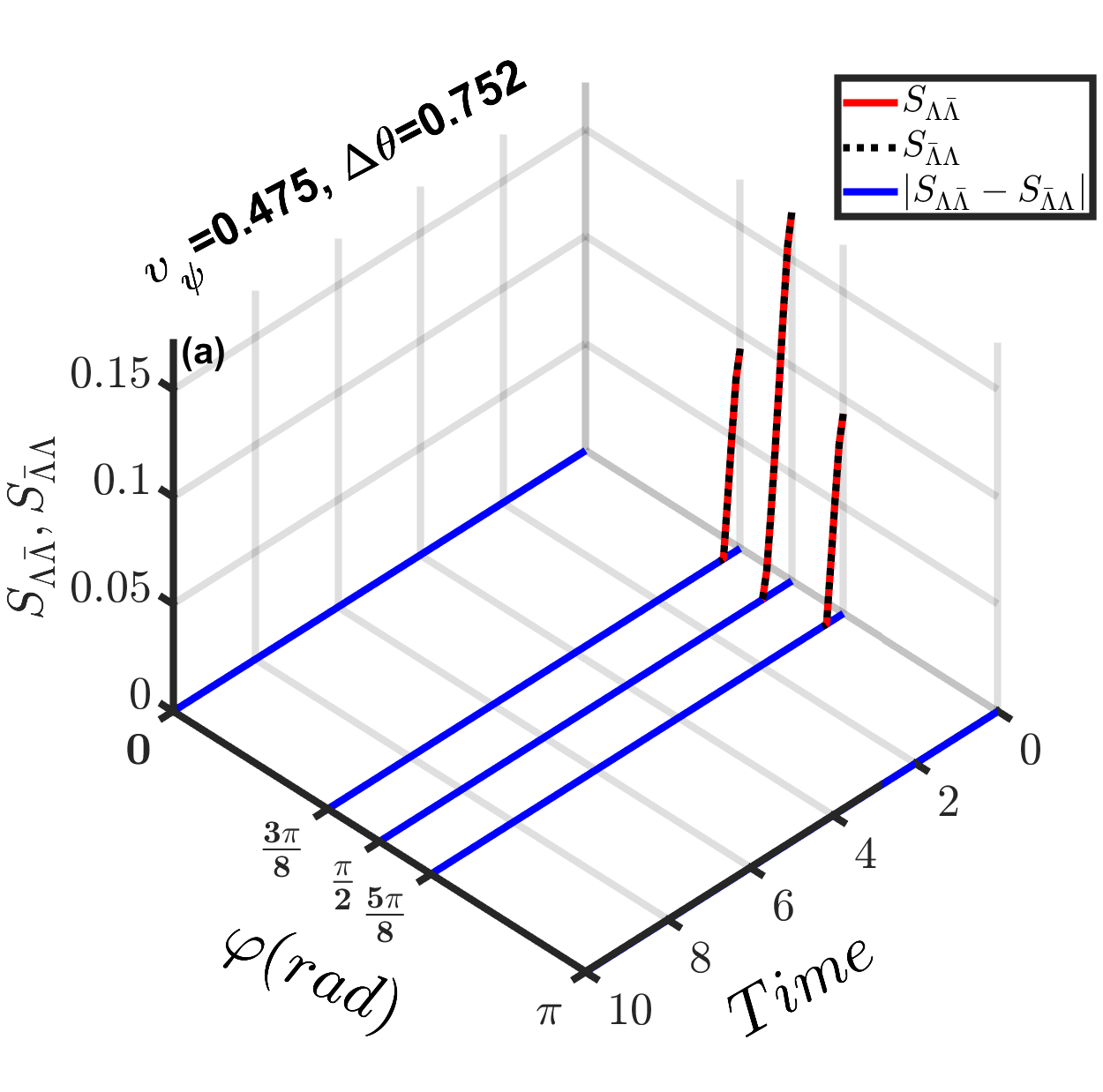}
\includegraphics[scale=0.4]{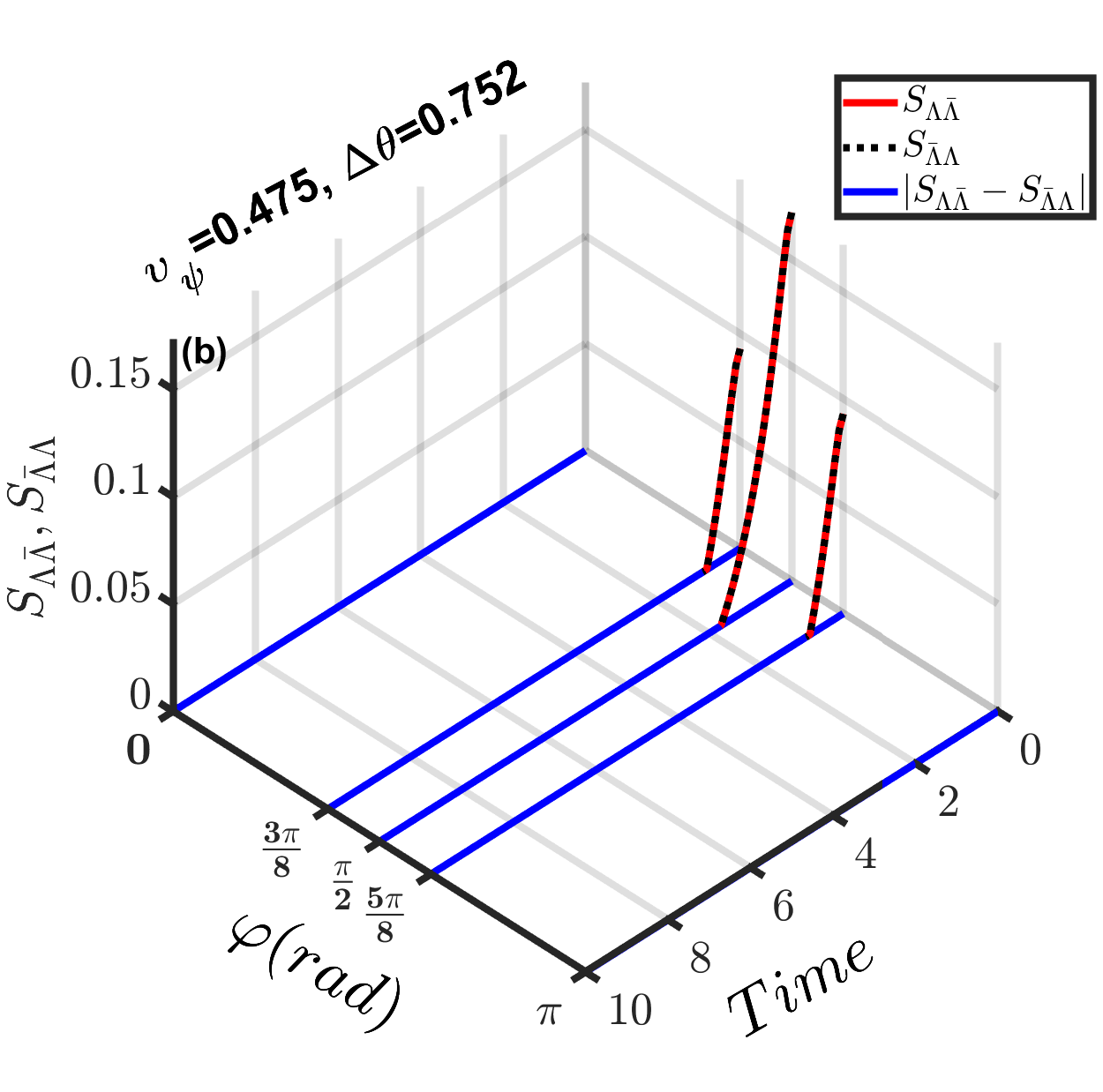}
\caption{Comparison of the quantum steerabilities $\mathcal{S}_{\Lambda\bar{\Lambda}}$ and $\mathcal{S}_{\bar{\Lambda}\Lambda}$ in the Markovian regime, for different values of $\varphi$ with $\mu = 0.6$ (a), and $\mu = 0.8$ (b), with $\tau = 0.1$.}
\label{fig:SC1}
\end{figure}

\begin{figure}[!h]
\includegraphics[scale=0.4]{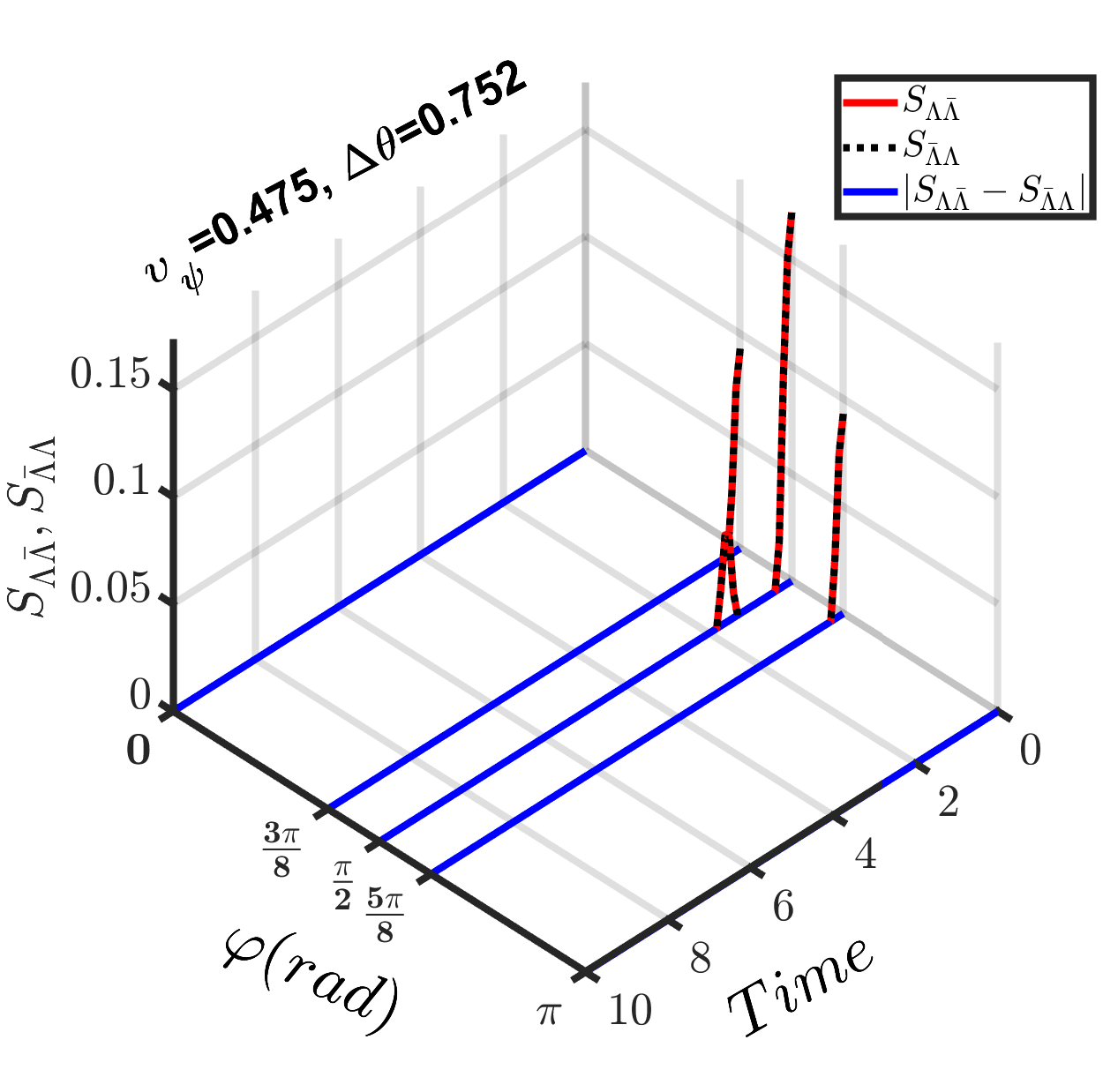}
\includegraphics[scale=0.4]{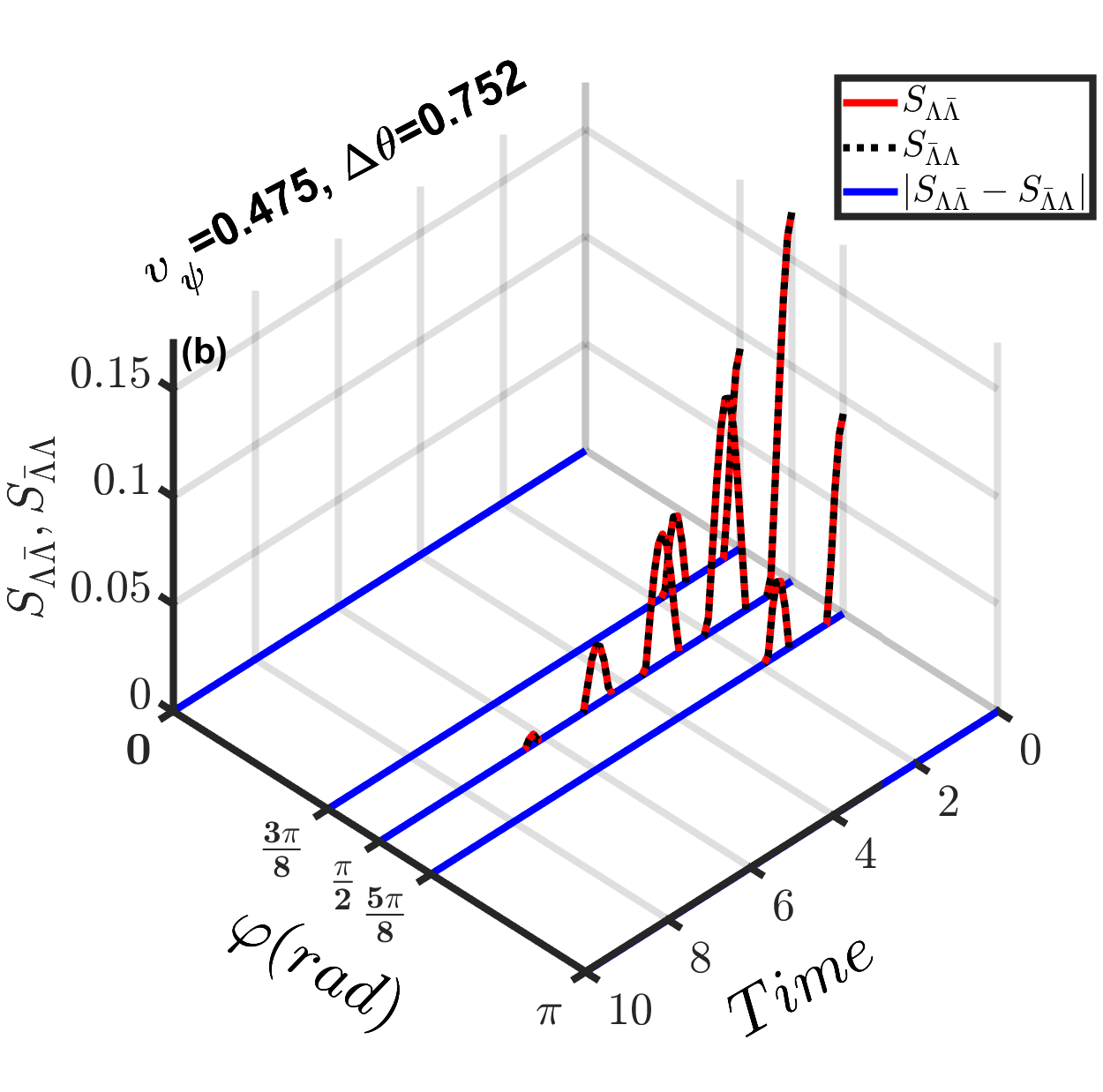}
\caption{Comparison of the quantum steerabilities $\mathcal{S}_{\Lambda\bar{\Lambda}}$ and $\mathcal{S}_{\bar{\Lambda}\Lambda}$ in the Non-Markovian regime, for various value of $\varphi$ with $\mu = 0.6$ (a), and $\mu = 0.8$ (b), with $\tau = 5$.}
\label{fig:SC2}
\end{figure}

We plot in Fig. \ref{fig:SC1}(a)-(b) the steerabilities $\mathcal{S}_{\Lambda\bar{\Lambda}}$ and $\mathcal{S}_{\Lambda\bar{\Lambda}}$, as well as the asymmetry $\delta \mathcal{S}$ versus the time for different values of the scattering angle $\varphi$ in the Markovian regime. We remark that quantum steerability can be enhanced by increasing classical correlations (quantified by $\mu$) in hyperon-antihyperon states. Furthermore, the steerabilities decrease quickly over time to vanish. For instance, for $\mu = 0.8$ and $\varphi = \pi/2$, the steerabilities $\mathcal{S}_{\Lambda\bar{\Lambda}} = \mathcal{S}_{\Lambda\bar{\Lambda}} > 0$ (i.e., $\delta \mathcal{S} = 0$) when time $< 0.8$. This indicates the existence of two-way steering between hyperon $\Lambda$ and antihyperon $\bar{\Lambda}$. However, for time $ > 0.8$, the steerability is vanishing, i.e., $\mathcal{S}_{\Lambda\bar{\Lambda}} = \mathcal{S}_{\Lambda\bar{\Lambda}} = 0$. This shows that there is no-way steering between the two qubits $\Lambda$ and $\bar{\Lambda}$.

Figures \ref{fig:SC2}(a)-(b) examine the non-Markovian regime, revealing oscillatory behavior in all quantities for lower $\mu$, which is attributed to memory effects. We note that the oscillations gradually lose amplitude over time. Similarly to the Markovian case, when $\mathcal{S}_{\Lambda\bar{\Lambda}} = \mathcal{S}_{\bar{\Lambda}\Lambda} > 0$ (i.e., $\delta \mathcal{S}= 0$). This indicates the presence of two-way steering between qubit $\Lambda$ and qubit $\bar{\Lambda}$. However, when $\mathcal{S}_{\Lambda\bar{\Lambda}} = \mathcal{S}_{\bar{\Lambda}\Lambda} = 0$, it shows that no-way steering exists between the two qubits.
\subsection{Entanglement of formation}
The entanglement of formation, $\mathcal{E}$, of a bipartite pure state is given by the von Neumann entropy of either of its subsystems, $\Lambda$ or $\bar{\Lambda}$. These entropies are equal and write as
\begin{equation}
\mathcal{E}(\hat\varrho) := -{\rm Tr}\Big\{\hat\varrho_{\Lambda} \log_2 \hat\varrho_{\Lambda}\Big\} = -\text{Tr}\Big\{\hat\varrho_{\bar{\Lambda}} \log_{2}\hat\varrho_{\bar{\Lambda}}\Big\},
\label{eq:EF}
\end{equation}
Here, $\hat\varrho_{\Lambda}$ is the partial trace of $\hat\varrho_{\Lambda\bar{\Lambda}}$ over subsystem $\bar{\Lambda}$, and $\hat\varrho_{\bar{\Lambda}}$ is defined similarly.\\
As shown in \citep{e4}, the entanglement defined in Eq. (\ref{eq:EF}) can be written as
\begin{equation}
\mathcal{E}(\hat{\varrho})=\mathbb{E}\Big(\mathcal{\rm C}(\hat{\varrho})\Big)
\end{equation}
with $\mathbb{E}\Big(\mathcal{\rm C}(\hat{\varrho})\Big)$ given by
\begin{equation}
\mathbb{E}\Big(\mathcal{\rm C}(\hat{\varrho})\Big) = g\left(\frac{1 + \sqrt{1-\mathcal{\rm C}^2(\hat{\varrho})}}{2}\right)
\end{equation}

where $g(x) = - x\log_2 x - (1 - x) \log_2 (1 - x)$. Such that $\mathcal{\rm C}$ represents the concurrence. It is essential to underline that the concurrence range varies between zero to one. The zero and nonzero values of concurrence indicate, respectively, a separable state and an entangled state.

The concurrence of a two-qubit mixed state $\hat{\varrho}_{\Lambda\bar{\Lambda}}$ is defined as the minimum average concurrence over all possible pure-state decompositions of $\hat{\varrho}_{\Lambda\bar{\Lambda}}$ \citep{e1}

\begin{equation}
\mathcal{\rm C}(\hat{\varrho}_{\Lambda\bar{\Lambda}})=\max\left\lbrace  \sqrt{\xi_{1}}-\sqrt{\xi_{2}}-\sqrt{\xi_{3}}-\sqrt{\xi_{4}},0\right\rbrace,
\label{eq:C}
\end{equation}

where $\xi_{i}$ are the eigenvalues of $\hat{\varrho}_{\Lambda\bar{\Lambda}}$, arranged in decreasing order

\begin{equation}
\mathcal{L}=\hat{\varrho}_{\Lambda\bar{\Lambda}}(\sigma_{y}\otimes\sigma_{y})\hat{\varrho}_{\Lambda\bar{\Lambda}}^{*}(\sigma_{y}\otimes\sigma_{y}),
\end{equation}
where $\hat{\varrho}^{*}_{\Lambda\bar{\Lambda}}$ denotes the complex conjugation of $\hat{\varrho}_{\Lambda\bar{\Lambda}}$ and $\sigma_{y}$ is the $y$-component of Pauli matrices. Nevertheless, for our purposes in this work, the concurrence of an X-state can be formulated by

\begin{equation}\label{crx}
\mathcal{\rm C}(\hat{\varrho}_{\Lambda\bar{\Lambda}})=2\max \left\lbrace |\eta\hat\varrho_{2,3}|-\sqrt{\hat\varrho_{1,1}\hat\varrho_{4,4}},|\eta\hat\varrho_{1,4}|-\sqrt{\hat\varrho_{2,2}\hat\varrho_{3,3}},0\right\rbrace .
\end{equation}
The concurrence for the hyperon-antihyperon system is found, using Eqs. (\ref{eq:varrho}) and (\ref{crx}), to be
\begin{equation}
\begin{aligned}
\mathcal{\rm C}(\hat{\varrho}_{\Lambda\bar{\Lambda}}) &= |\eta \gamma_{2}| \\
&= \frac{\bigg|\eta\left(1 + \upsilon_{\psi} - \sqrt{\left(1 + \upsilon_{\psi} \cos(2\varphi)\right)^2 - \big( 1- \upsilon^2_{\psi}\big)\sin^{2}(\Delta\theta) \sin^2(2\varphi)}\right)\bigg|}{2\big(1 + \upsilon_{\psi} \cos^2(\varphi)\big)}
\end{aligned}
\end{equation}

\begin{figure*}[t]
\includegraphics[scale=0.45]{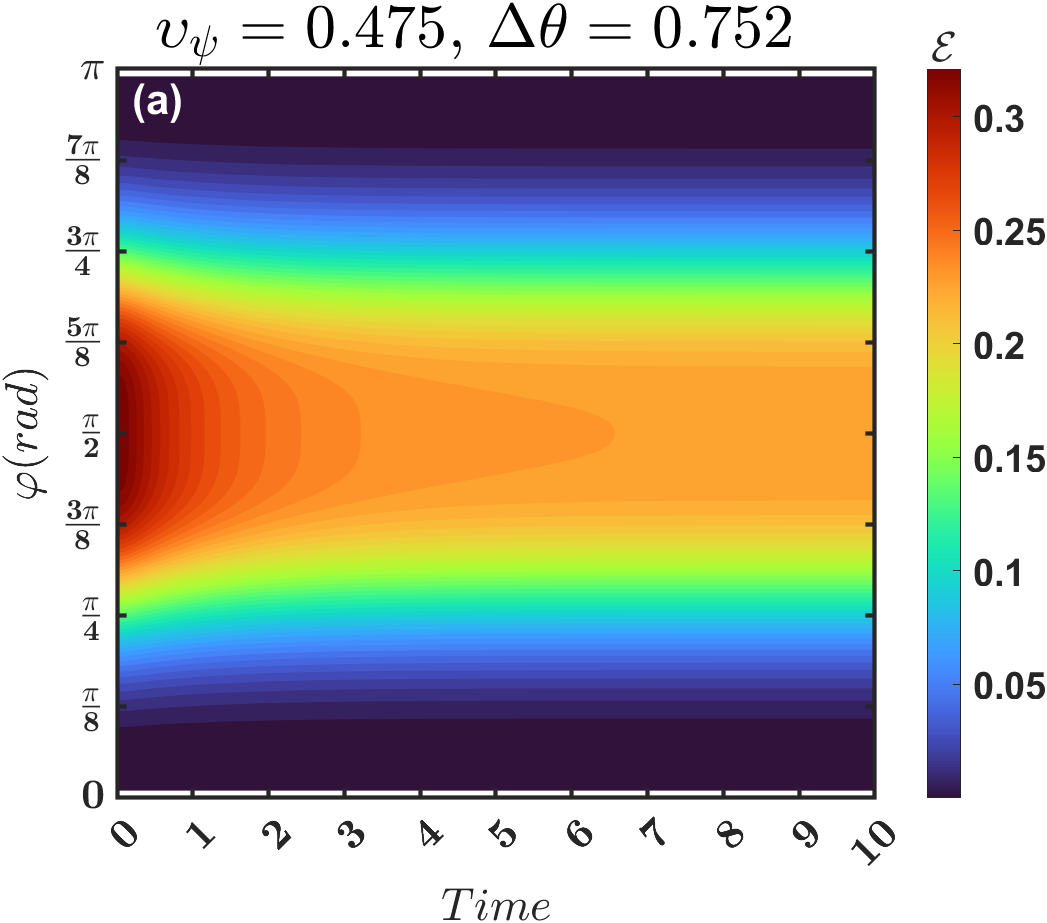}
\includegraphics[scale=0.45]{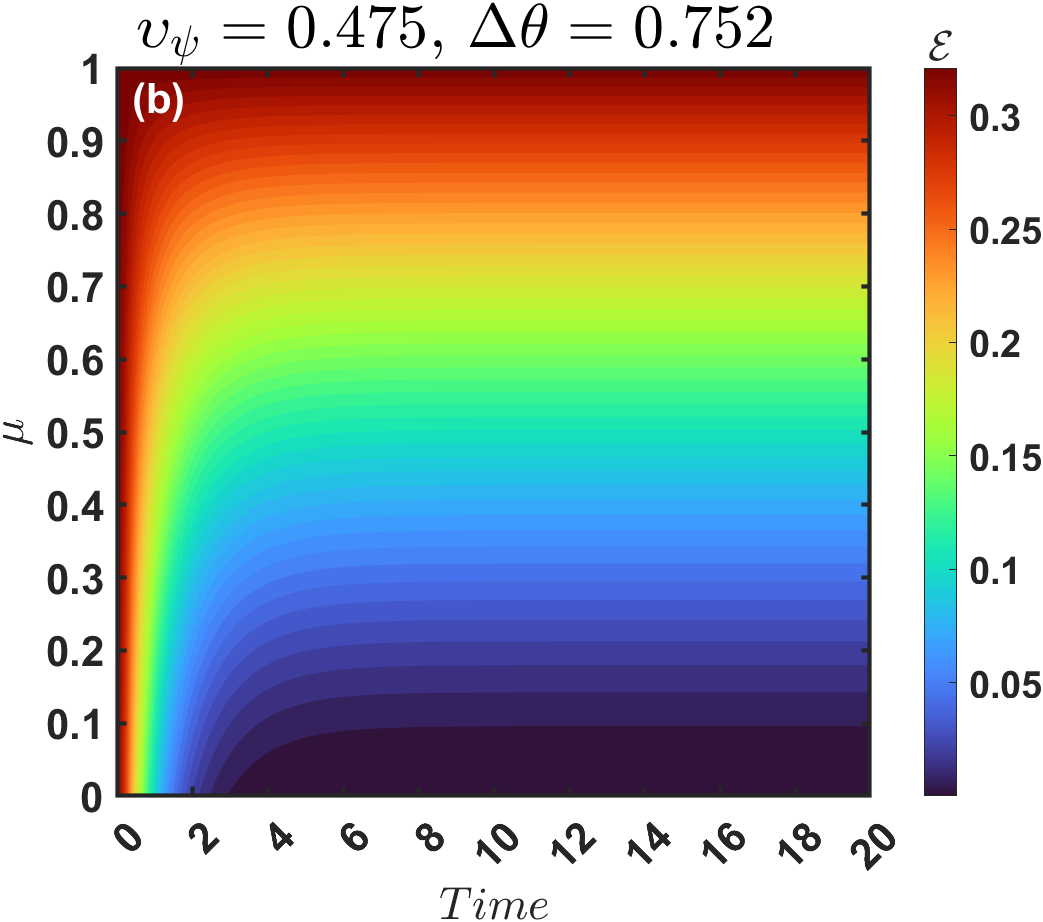}
\caption{Dynamical evolution of entanglement of formation $\mathcal{E}$ in the Markovian regime with $\tau = 0.1$ (a) $\mu = 0.8$ and (b) $\varphi=\pi/2$.}
\label{fig:e1}
\end{figure*}

Fig. \ref{fig:e1}(a) shows the evolution of entanglement of formation $\mathcal{E}$ as a function of time and the scattering angle $\varphi $ in Markovian regime. We observe that entanglement of formation gradually decreases over time to be stationary approximately at $\mathcal{E}=0.22$. The maximum of the entanglement is achieved for the angle around $\varphi= \pi/2$, as depicted in Fig. \ref{fig:e1}(a). Also, the entanglement exhibits remarkable symmetry around $\varphi = \pi/2$. We notice that the entanglement decreases as $\varphi$ deviates further from $\pi/2$. 

In Fig. \ref{fig:e1}(b), we present $\mathcal{E}$ as a function of time and $\mu$ in Markovian regime. It is observed that stronger classical correlations effectively delay entanglement decay, thereby reducing decoherence. Entanglement reaches a steady state after a time threshold (Fig. \ref{fig:e1}(b)). Steady-state entanglement grows as $\mu$ increases. 

\begin{figure*}[t]
\includegraphics[scale=0.45]{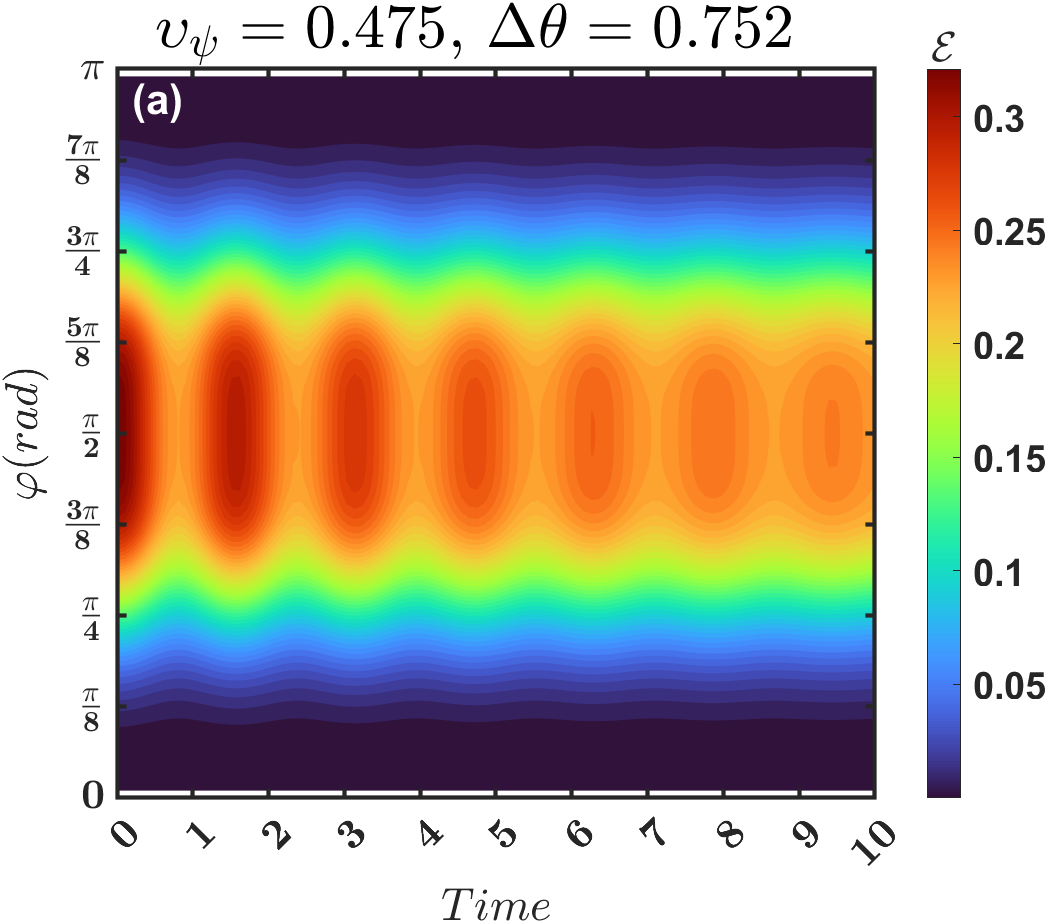}
\includegraphics[scale=0.45]{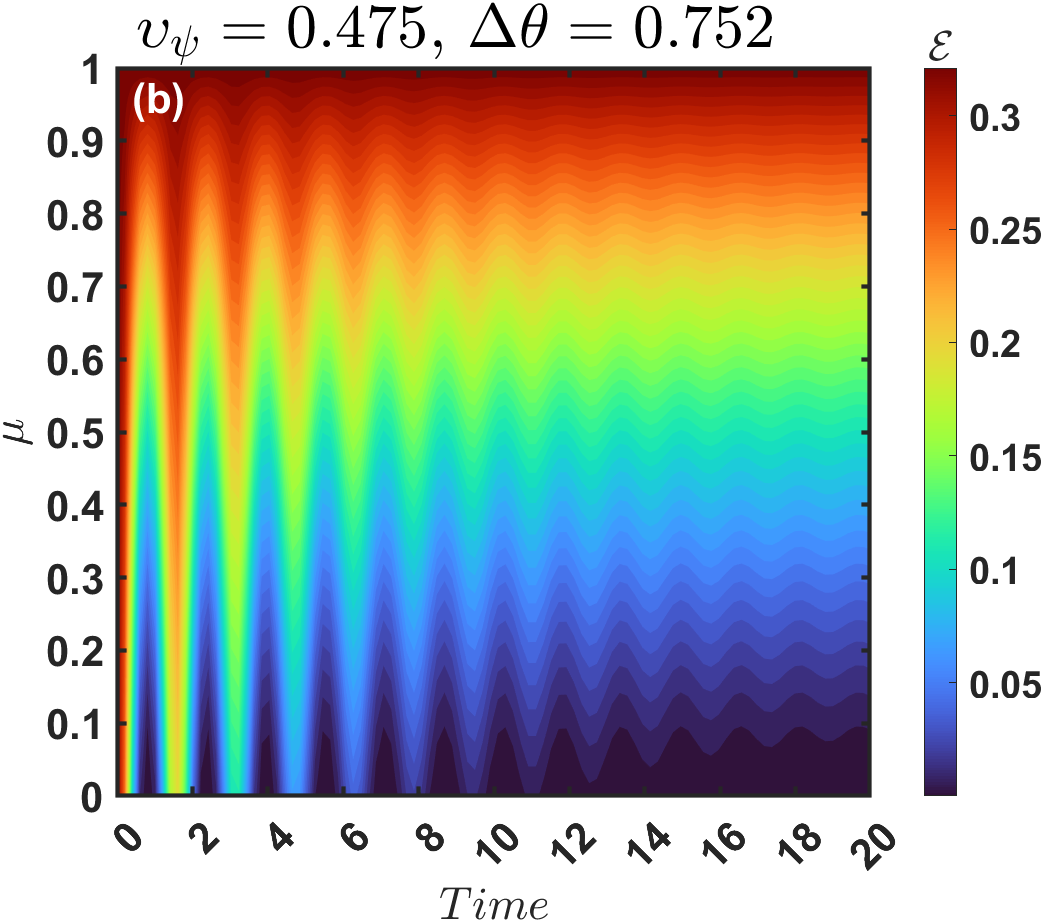}
\caption{Dynamical evolution of entanglement of formation $\mathcal{E}$ in the Non-Markovian regime with $\tau = 5$ (a) $\mu = 0.8$ and (b) $\varphi=\pi/2$.}
\label{fig:e2}
\end{figure*}

In Fig. \ref{fig:e2}(a), we observe that entanglement of formation $\mathcal{E}$ does not follow a monotonic decay over time. On the contrary, it exhibits periodic oscillations. More specifically, entanglement \(\mathcal{E}\) reaches its maximum at $\varphi=\pi/2$. However, memory effects lead to a non-monotonic behavior of entanglement of formation over time.

Fig. \ref{fig:e2}(b) shows that the entanglement of formation $\mathcal{E}$ oscillations decreases as the classical correlation parameter $\mu$ increases. Moreover, the decay rate becomes significantly slower for higher values of $\mu$. This relationship highlights the delicate balance between classical and quantum correlations, emphasizing how classical parameters can influence quantum dynamics.
\subsection{Geometric quantum discord}
The quantum discord provides a more comprehensive measure of quantum correlations in a system, taking into account all forms of quantum correlation. It was initially introduced as an entropy-based measure that assesses the actual quantum correlations present in a quantum state \citep{ref12}. According to this measure, the mutual information between subsystems $\Lambda$ and $\bar{\Lambda}$ is represented by 
\begin{equation}
{\rm Q}[\hat{\varrho}_{\Lambda\bar{\Lambda}}] ={\rm I}(\hat{\varrho}_{\Lambda}:\hat{\varrho}_{\bar{\Lambda}})-\mathcal{\rm C}(\hat{\varrho}_{\Lambda\bar{\Lambda}}),
\end{equation}
where ${\rm I}(\hat{\varrho}_{\Lambda}:\hat{\varrho}_{\bar{\Lambda}})=S(\hat{\varrho}_{\Lambda})+S(\hat{\varrho}_{\bar{\Lambda}})-S(\hat{\varrho}_{\Lambda\bar{\Lambda}})$ represents the mutual information between subsystems $\Lambda$ and $\bar{\Lambda}$, while $\mathcal{\rm C}(\hat{\varrho}_{\Lambda\bar{\Lambda}})$ denotes the classical correlation of the composite system $\hat{\varrho}_{\Lambda\bar{\Lambda}}$, defined as $\mathcal{\rm C}(\hat{\varrho}_{\Lambda\bar{\Lambda}})=\max_{\bar{\Lambda}_{k}}[S(\hat{\varrho}_{\Lambda}-\sum_{k} p_{k}S(\hat{\varrho}_{k})]$. This involves maximizing over positive operator-valued measurements (POVMs), $\bar{\Lambda}_{k}$, applied exclusively to subsystem $\bar{\Lambda}$. However, even for a two-qubit system \citep{d2,ref13}, performing this analytical maximization over POVMs is a complex task. Faced with this challenge, various alternative approaches have been developed to assess quantum correlations \citep{d5}. These methods incorporate geometric considerations rather than relying solely on entropic calculations \citep{d7}, thus providing alternative options to characterize the quantum correlations present in quantum systems.\\
Geometric approaches are widely employed to assess and quantify quantum resources across various quantum systems, with particular emphasis on the Schatten 1-norm (or trace norm) quantum discord \citep{ref12}. The GQD, for a two-qubit state, can be expressed as \citep{d10}
\begin{equation}
\mathcal{D_{G}}(\hat{\varrho}_{\Lambda\bar{\Lambda}})=\min_{\hat{\varrho}_{c} \in \Omega} ||\hat{\varrho}_{\Lambda\bar{\Lambda}} - \hat{\varrho}_{c}||_1,
\label{eq:QG}
\end{equation}
where  $||\hat{\varrho}_{\Lambda\bar{\Lambda}} - \hat{\varrho}_{c}||_1=\textmd{tr}\sqrt{(\hat{\varrho}_{\Lambda\bar{\Lambda}} - \hat{\varrho}_{c})^{\dagger}(\hat{\varrho}_{\Lambda\bar{\Lambda}} - \hat{\varrho}_{c})}$ represents the Schatten 1-norm. The set $\Omega$ of closest classical-quantum states $\hat{\varrho}_{c}$ with respect to local measurements
on subsystem $A$ is given by
\begin{equation}
\hat{\varrho}_{c} = \sum_{k} p_{k} \Pi_{k,\Lambda} \otimes \hat{\varrho}_{k,\bar{\Lambda}},
\end{equation}
where $p_{k}\in [0,1] $ and $\sum_{k} p_{k}=1$, in which $\left\lbrace p_{k}\right\rbrace$ is a probability distribution. The orthogonal projectors associated with qubit $A$ are denoted by $\Pi_{k,\Lambda}$, and the density matrix associated with the second qubit is $\varrho_{k,\bar{\Lambda}}$.\\
The minimization solution in Eq. (\ref{eq:QG}) for a generic two-qubit X state enables us to express the GQD based on the Schatten 1-norm as \cite{d10}
\begin{equation}
\mathcal{D_{G}}(\hat{\varrho}_{\Lambda\bar{\Lambda}})=\frac{1}{2}\sqrt{\frac{R_{1,1}^{2}R_{\max}^2 -R_{2,2}^{2}R_{\min}^2}{R_{\max}^2-R_{\min}^2 +R_{1,1}^{2}-R_{2,2}^{2}}},
\label{eq:Q}
\end{equation}
where $$R_{\max}^2=\max\left\lbrace R_{2,2}^{2}+R_{3,0}^{2},R_{3,3}^{2}\right\rbrace$$
and
$$R_{\min}^2=\min\left\lbrace R_{1,1}^{2},R_{3,3}^{2}\right\rbrace.$$

Besides, $R_{\alpha,\beta}$ in Eq. (\ref{eq:Q}) being the components of the correlation matrix occurring after decomposing the state $R$ in the Fano-Bloch representation as
\begin{equation}
R=\frac{1}{4}\sum_{\alpha,\beta=0}^{3}R_{\alpha,\beta}\sigma_{\alpha}\otimes\sigma_{\beta},
\end{equation}
where the non vanishing matrix elements $R_{\alpha,\beta}$ are given by

$$
\begin{aligned}
\phantom{h(x) }R_{1,1} &=\textmd{tr}\left[\big(\sigma_{1}\otimes\sigma_{1}\big)\hat{\varrho}_{\Lambda\bar{\Lambda}}\right] = 2\eta\left(\hat{\varrho}_{2,3}+\hat{\varrho}_{1,4}\right) ,\\
\phantom{h(x) }R_{2,2} &=\textmd{tr}\left[\big(\sigma_{2}\otimes\sigma_{2}\big)\hat{\varrho}_{\Lambda\bar{\Lambda}}\right]=2\eta\left(\hat{\varrho}_{2,3}-\hat{\varrho}_{1,4}\right),\\
\phantom{h(x)}R_{3,3} &=\textmd{tr}\left[\big(\sigma_{3}\otimes\sigma_{3}\big)\hat{\varrho}_{\Lambda\bar{\Lambda}}\right]=1-2\left(\hat{\varrho}_{2,2}+\hat{\varrho}_{3,3}\right),\\
\phantom{h(x)}R_{0,3} &=\textmd{tr}\left[\big(\sigma_{0}\otimes\sigma_{3}\big)\hat{\varrho}_{\Lambda\bar{\Lambda}}\right]=2\left(\hat{\varrho}_{1,1}\textcolor{red}{+}\hat{\varrho}_{2,2}\right) -1,\\
R_{3,0} &=\textmd{tr}\left[\big(\sigma_{0}\otimes\sigma_{3}\big)\hat{\varrho}_{\Lambda\bar{\Lambda}}\right]=2\left(\hat{\varrho}_{1,1}\textcolor{red}{+}\hat{\varrho}_{2,2}\right) -1.
\end{aligned}
$$

\begin{figure}[t]
\includegraphics[scale=0.45]{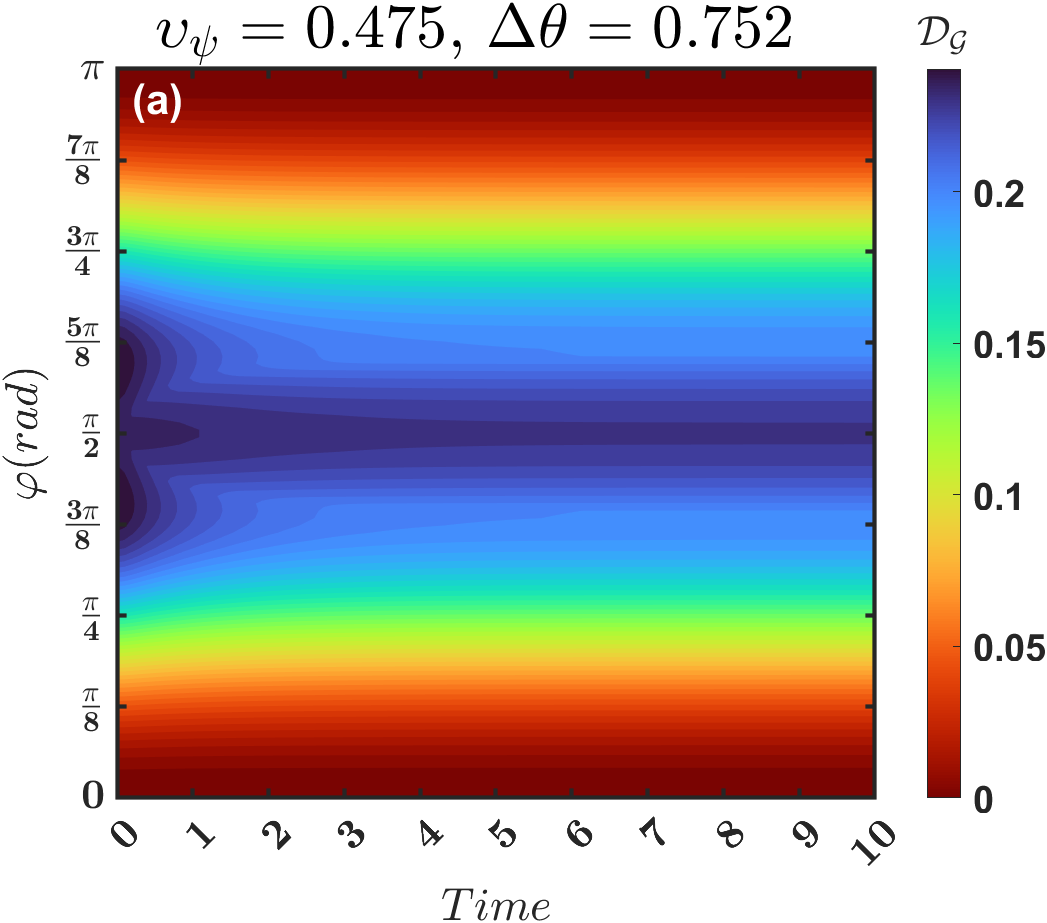}
\includegraphics[scale=0.45]{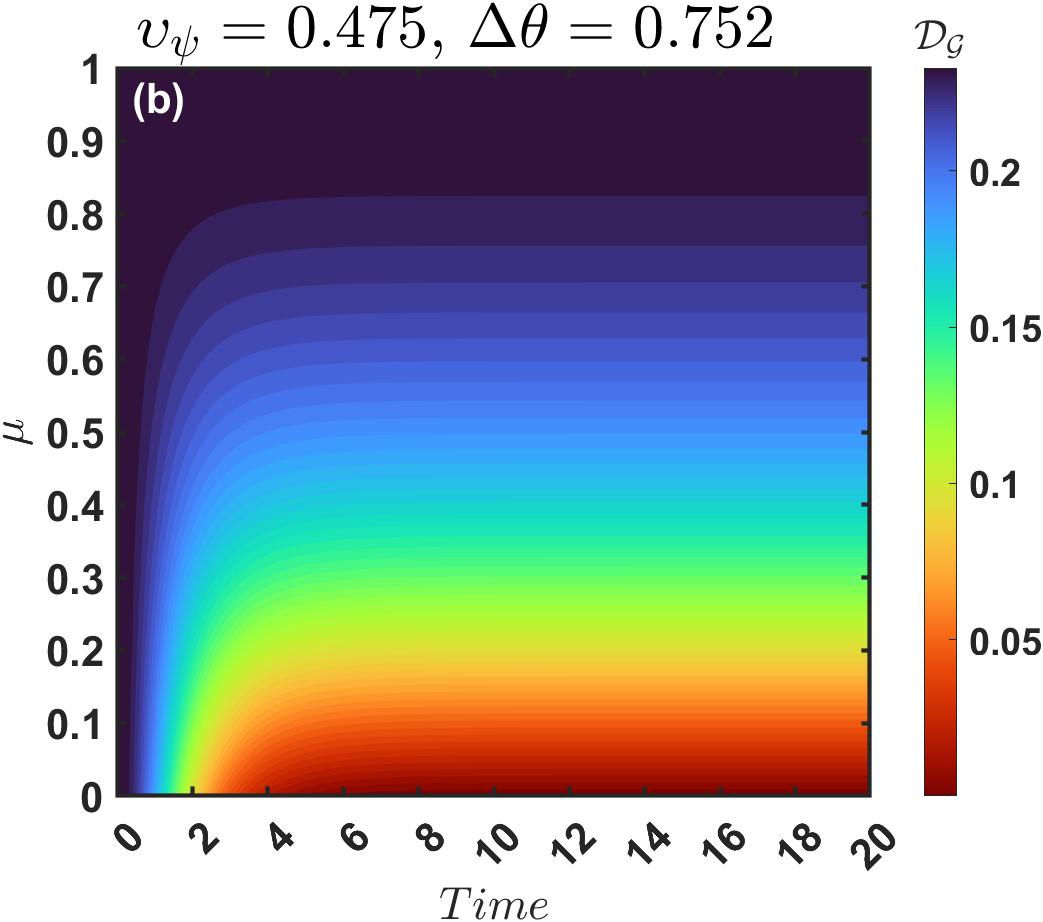}
\caption{Dynamical evolution of GQD $\mathcal{D_{G}}$ in the Markovian regime with $\tau =0.1$ for (a) $\mu = 0.8$ and (b) $\varphi=\pi/2$.}
\label{fig:d1}
\end{figure}
\begin{figure}[t]
\includegraphics[scale=0.45]{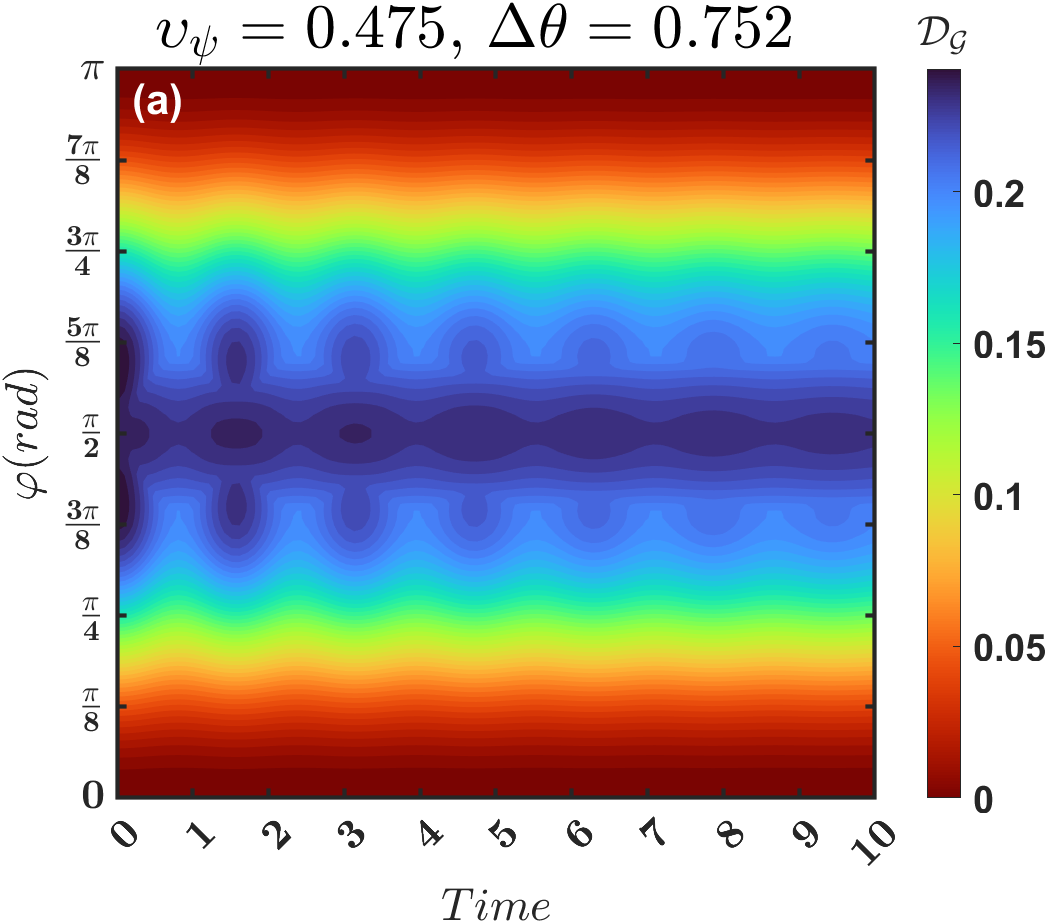}
\includegraphics[scale=0.45]{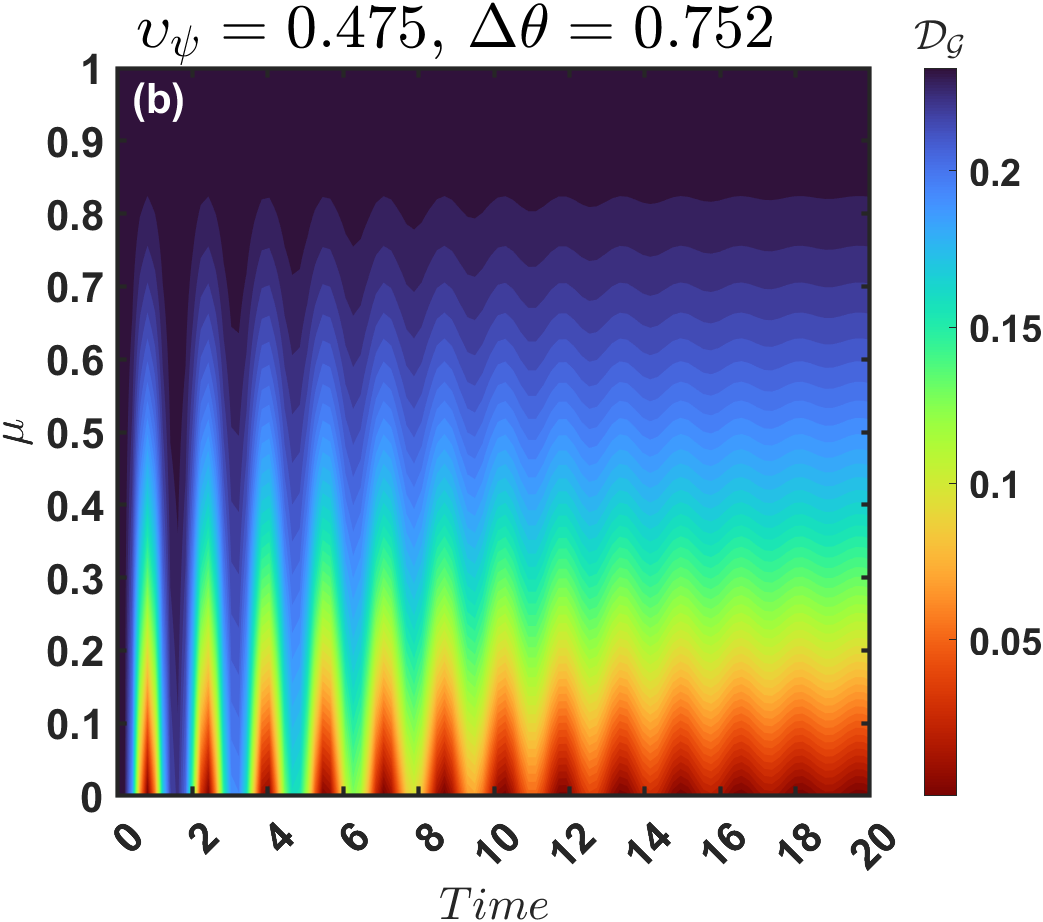}
\caption{Dynamical evolution of GQD $\mathcal{D_{G}}$ in the Non-Markovian regime with $\tau =5$ for (a) $\mu = 0.8$ and (b) $\varphi=\pi/2$.}
\label{fig:d2}
\end{figure}

The geometric quantum discord $\mathcal{D_{G}}$ is plotted in Fig. \ref{fig:d1} (Markovian regime) and Fig. \ref{fig:d2} (non-Markovian regime). The same parameter domain used for quantum steering and concurrence was employed for these GQD calculations.

Figure \ref{fig:d1} depicts the time evolution of GQD in the Markovian regime, with (a) $\varphi$ and (b) $\mu$ varied in separate subfigures. From Fig. \ref{fig:d1} (a), we see that GQD is symmetric and achieves its maximum value around \(\pi/2\). A gradual decrease in GQD is observed over time. 

In Fig. \ref{fig:d2} (a), we show GQD in the non-Markovian regime as a function of time and the scattering angle $\varphi$. It is observed that GQD does not follow a monotonic decay over time. On the contrary, it exhibits periodic oscillations, characteristic of non-Markovian effects. More specifically, GQD reaches a maximum at $\varphi=\pi/2$. However, memory effects result in a non-monotonic behavior of GQD over time. 

Figures \ref{fig:d1}(b) and \ref{fig:d2}(b) illustrate how classical correlations in the dephasing channel affect GQD evolution. Stronger classical correlations, indicated by higher values of $\mu$, lead to a slower decay of GQD, thus enhancing the preservation of quantum advantages. Figure \ref{fig:d2}(b) also exhibits this trend; however, in the non-Markovian regime, significant memory effects give rise to non-exponential decay and complex dynamics in GQD. For $\mu$ close to 1, GQD is robust against decoherence in both regimes.
\subsection{Quantum Coherence}

Quantum coherence is a vital resource for various quantum information tasks. The $l_{1}$-norm of coherence is both intuitive and computable, making it a useful measure of quantum coherence \citep{c1,c2}. The quantification of coherence by this measure relies on determining the minimal distance between the given quantum state and the set of all incoherent states. To calculate the $l_{1}$-norm coherence of $\hat{\varrho} =\sum_{i,j}\hat{\varrho}_{i,j}\ket{i}\bra{j}$, sum the absolute values of its off-diagonal elements
\begin{figure}[!h]
\includegraphics[scale=0.45]{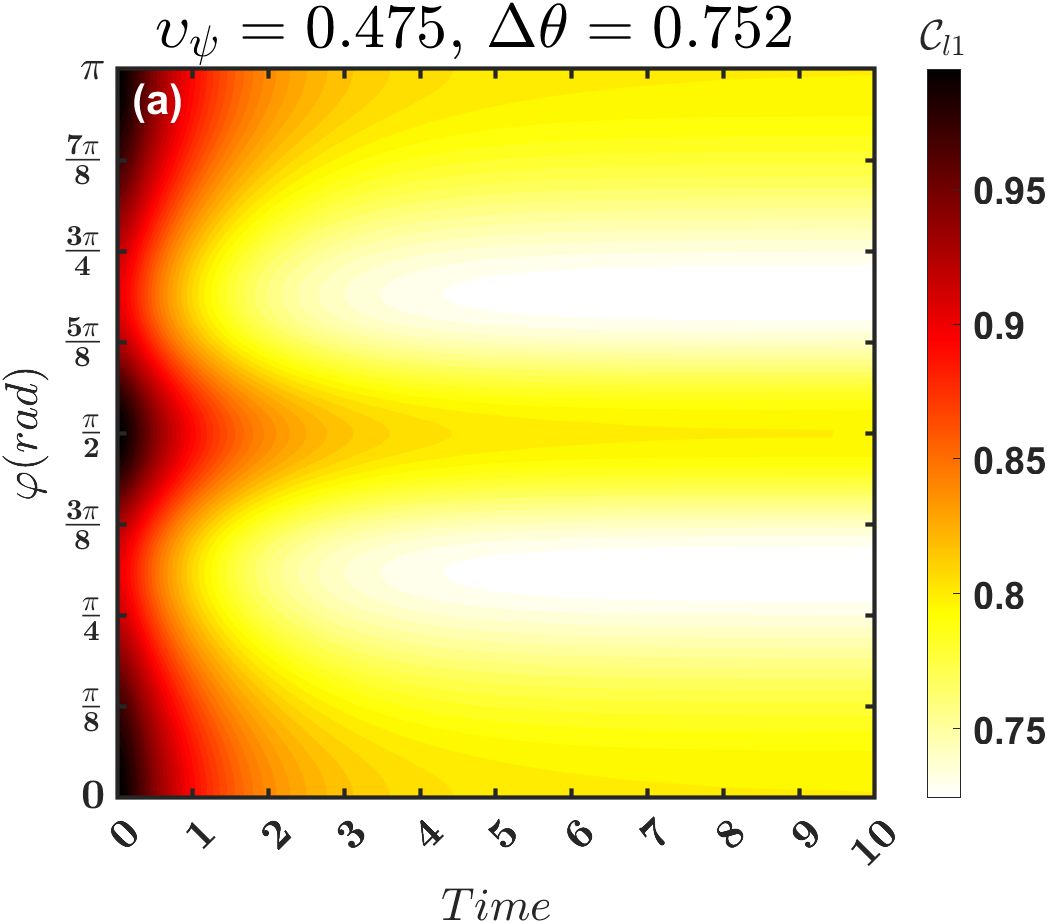}
\includegraphics[scale=0.45]{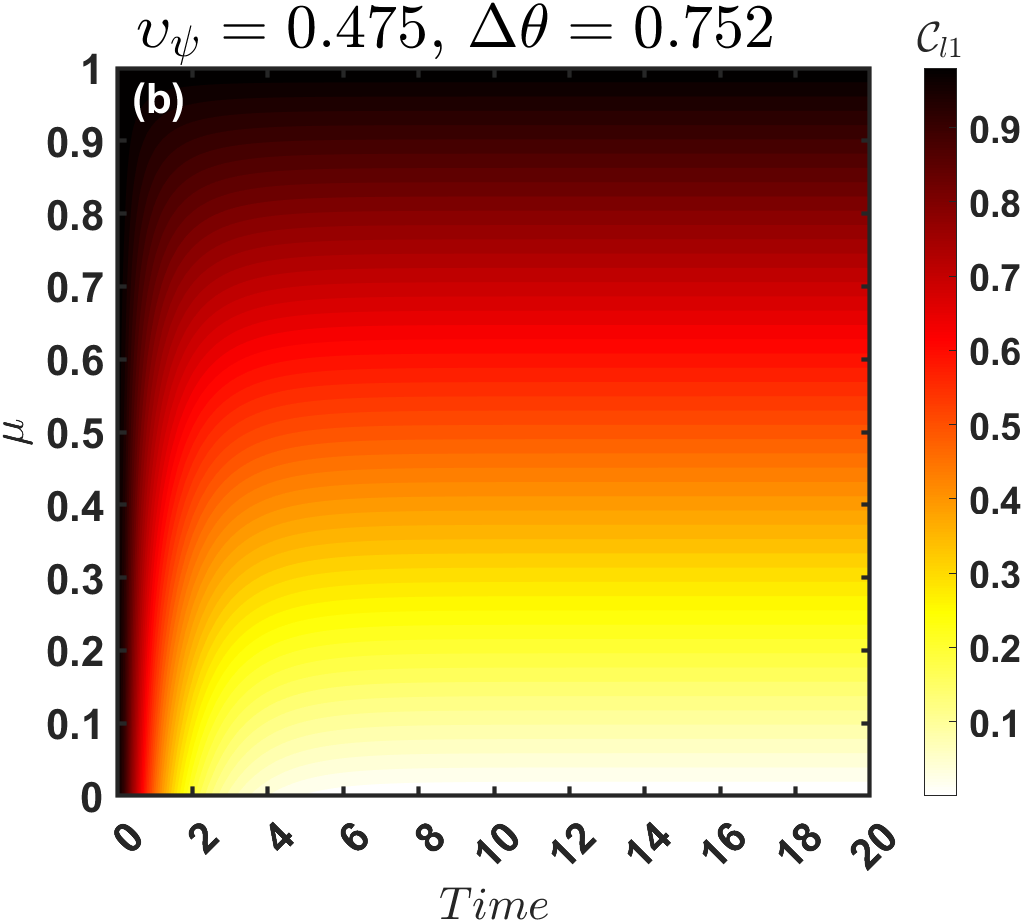}
\caption{Dynamical evolution of quantum coherence $\mathcal{C}_{l1}$ in the Markovian regime with $\tau = 0.1$ for (a) $\mu = 0.8$ and (b) $\varphi=\pi/2$.}
\label{fig:2}
\end{figure}
\begin{figure}[!h]
\includegraphics[scale=0.45]{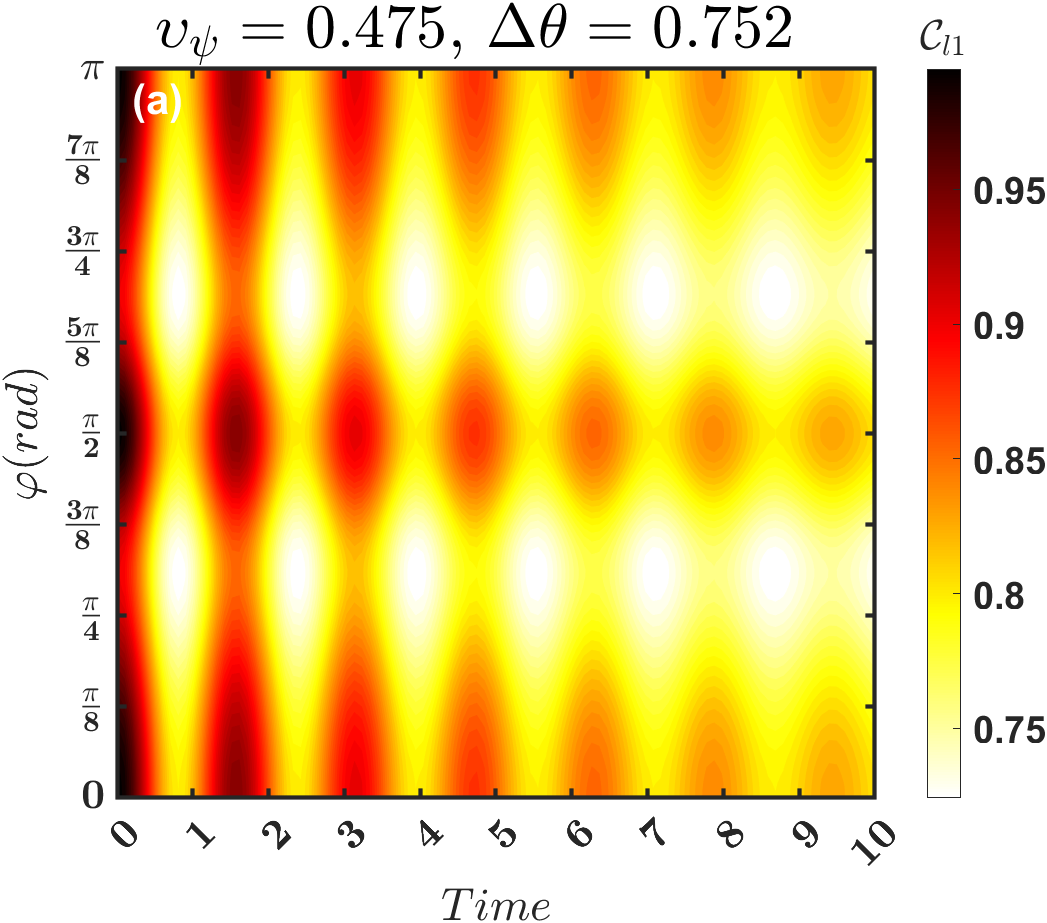}
\includegraphics[scale=0.45]{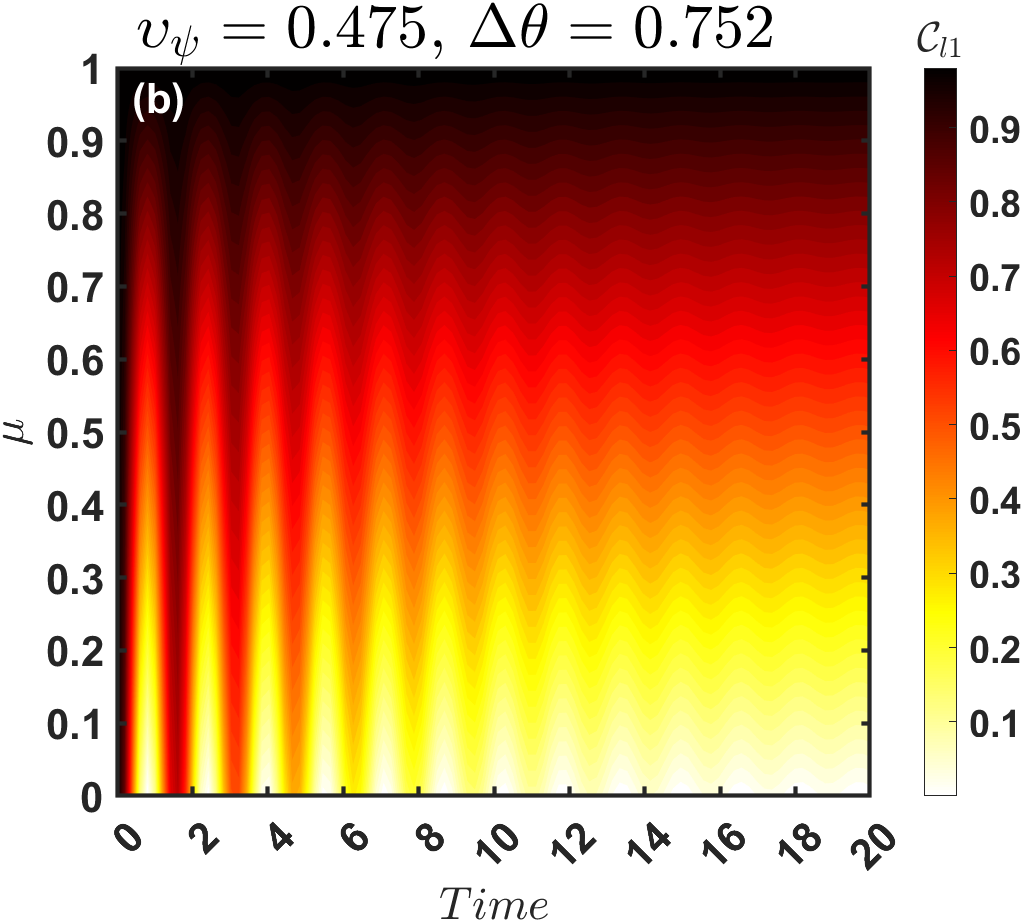}
\caption{Dynamical evolution of quantum coherence $\mathcal{C}_{l1}$ in the Non-Markovian regime with $\tau = 5$ for (a) $\mu = 0.8$ and (b) $\varphi=\pi/2$.}
\label{fig:3}
\end{figure}
\begin{equation}
\mathcal{C}_{l1}(\hat{\varrho}_{\Lambda\bar{\Lambda}})=\sum_{i\neq j}|\hat{\varrho}_{i,j}|
\end{equation}
From Eq. (\ref{eq:varrho}) the $l_{1}$-norm is given by
\begin{equation}
\mathcal{C}_{l1}=2|\eta\hat\varrho_{2,3}|+2|\eta\hat\varrho_{1,4}|
\label{eq:C0}
\end{equation}

Fig.~\ref{fig:2}(a) shows the evolution of \( \mathcal{C}_{l1} \) as a function of time and the angle $\varphi$. We observe that quantum coherence gradually decreases over time, reaching its maximum at specific angles: $\varphi = 0$, $\varphi = \pi$, and $\varphi=\pi/2$. These angles indicate that the alignment of particle spins plays a crucial role in maximizing quantum correlations. Additionally, coherence displays a remarkable symmetry around $\varphi =\pi/2$.

Coherence dynamics are affected by the stronger classical correlations that arise from increasing $\mu$. The behavior of $\mathcal{C}_{l1}$ as a function of time and $\mu$ is depicted in Fig. \ref{fig:2}(b). We remark as time progresses, quantum coherence increases and eventually reaches a constant value, indicating a steady state. Also, we observe a decrease in the coherence decay rate with increasing $\mu$. This indicates that classical correlations delay decoherence.

For the same time intervals, coherence evolution is shown in Fig. \ref{fig:2} (Markovian) and Fig. \ref{fig:3} (non-Markovian). In the non-Markovian regime, memory effects and non-exponential decay cause oscillations in coherence.

In Fig. \ref{fig:3}(a), we see that quantum coherence exhibits non-monotonic decay over time. Instead, it oscillates periodically, indicating non-Markovian behavior. This can explaine by the memory effects. Besides, the maximum values of quantum coherence are observed at $\varphi = 0$, $\varphi = \pi/2$, and $\varphi= \pi$.

Fig. \ref{fig:3} (b) shows that the amplitude of the quantum coherence oscillations decreases as the classical correlation parameter, $\mu$, increases. Slower decay at higher $\mu$ confirms that stronger classical correlations preserve coherence.
\section{Comparative analysis of quantum resources}\label{sec:4}
In this section, we discuss the influence of classical correlations on the temporal evolution of quantum resources in both the Markovian regime (memoryless dynamics) and the non-Markovian regime (memory effects are crucial).
\begin{figure}[!h]
    \includegraphics[scale=0.4]{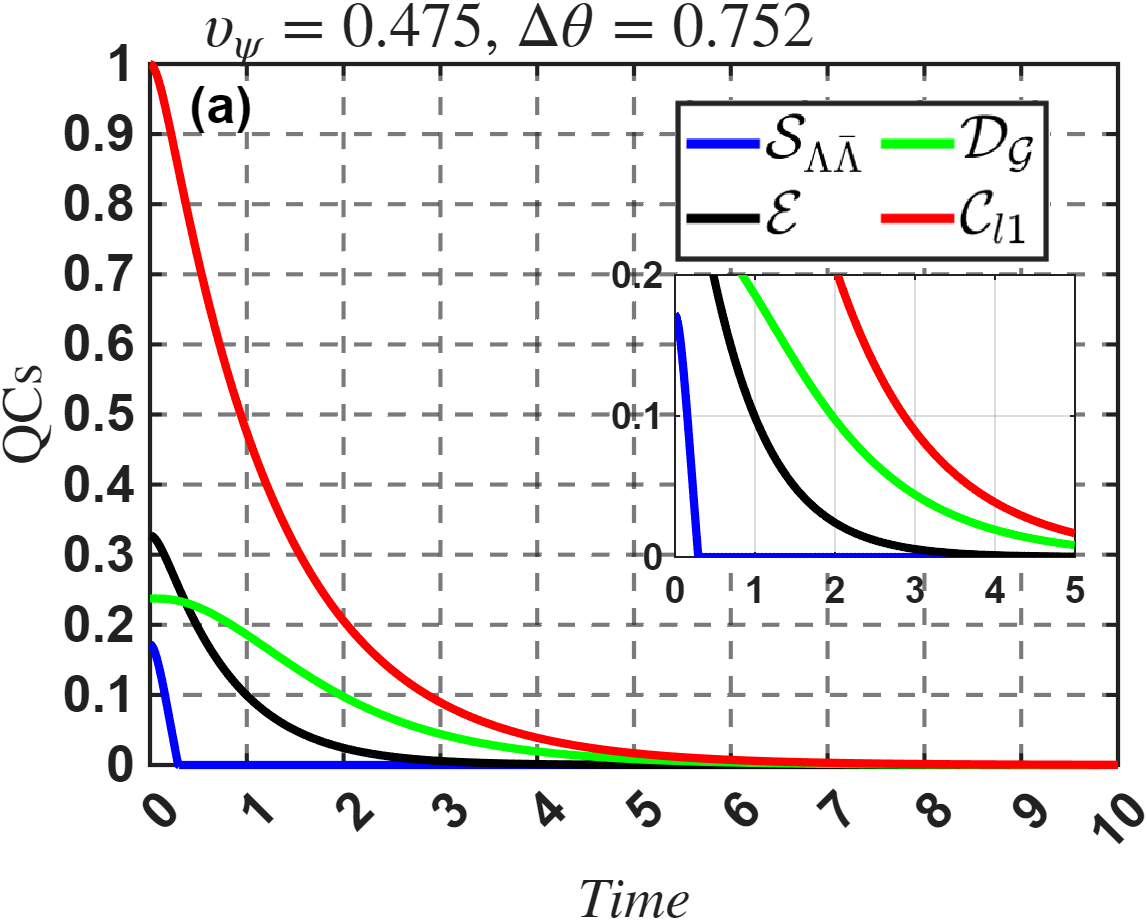}
    \includegraphics[scale=0.4]{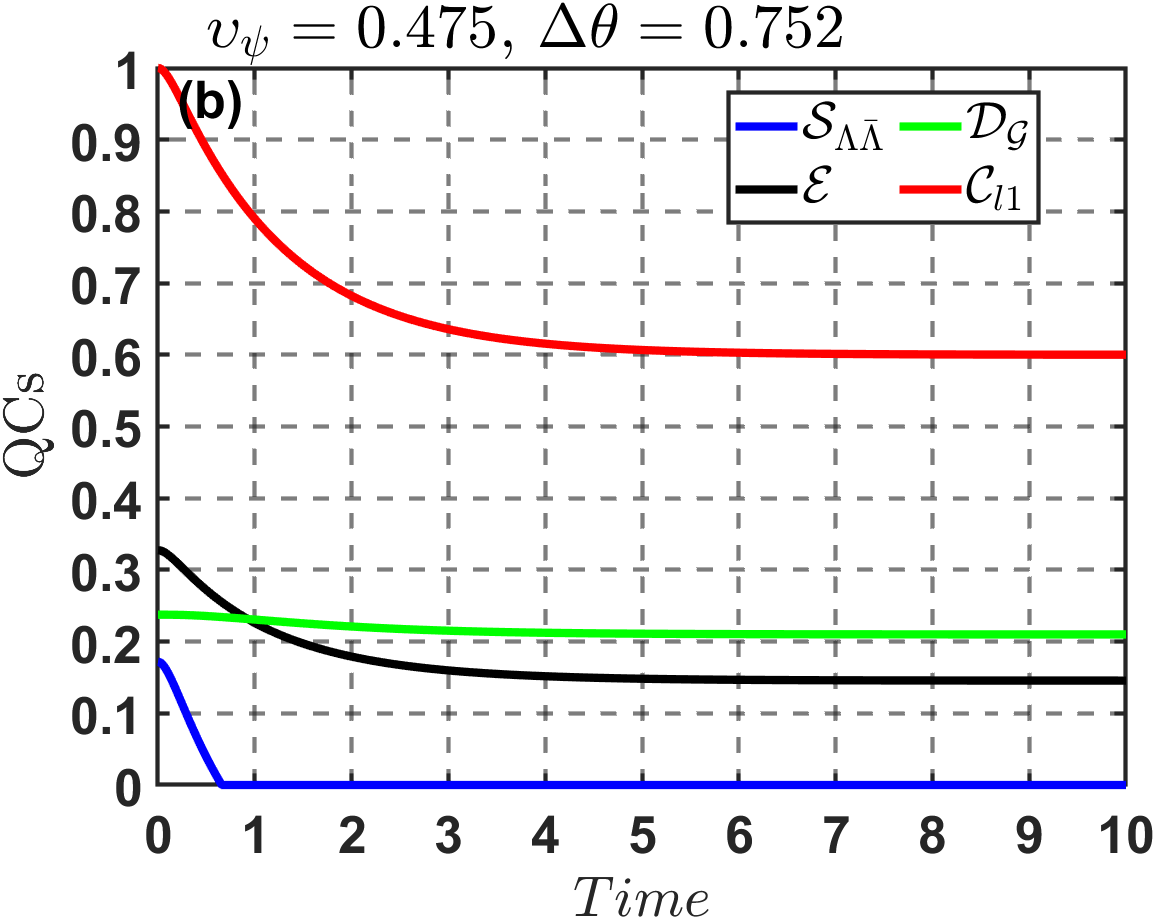}\\
     
    \includegraphics[scale=0.4]{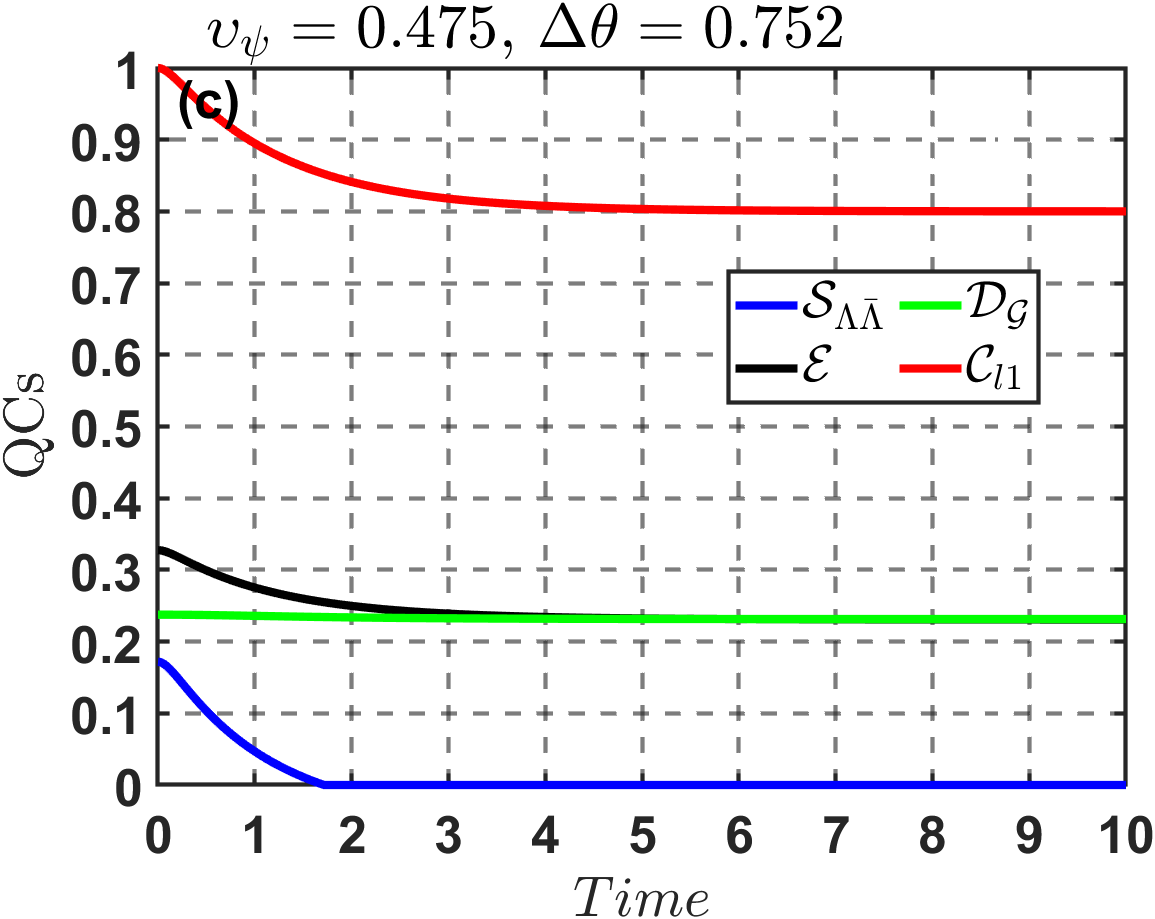}
    \includegraphics[scale=0.4]{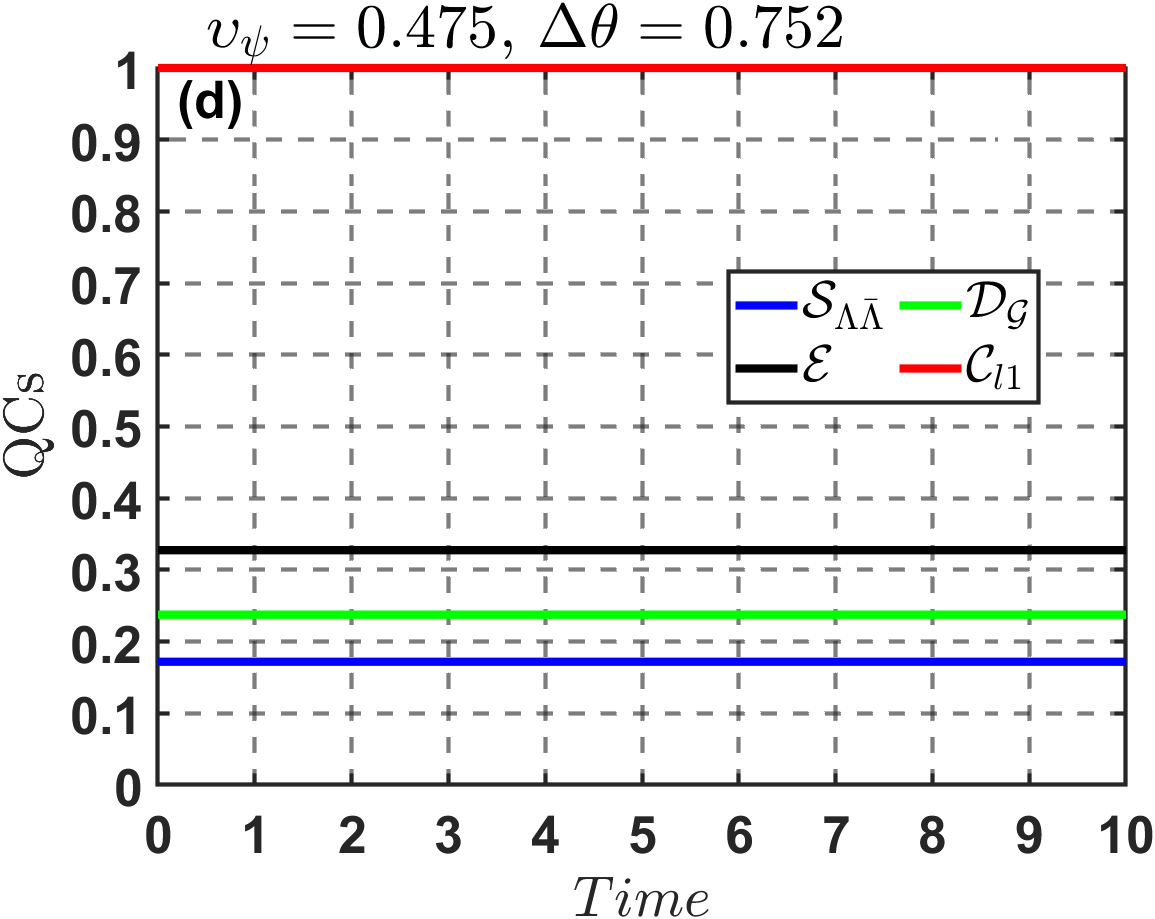}
\caption{Dynamical evolution of quantum steering $\mathcal{S}_{\Lambda\bar{\Lambda}}$, entanglement of formation \(\mathcal{E}\), GQD $\mathcal{D_{G}}$, and $l_1$-norm of quantum coherence $\mathcal{C}_{l1}$ in the Markovian regime for different values of $\mu$ : $\mu=0$ (a), $\mu=0.6$ (b), $\mu=0.8$ (c) and $\mu=1$ (d). With $\tau = 0.1$ for $\varphi = \pi/2$.}
\label{fig:M}
\end{figure}

\begin{figure}[!h]
    \includegraphics[scale=0.4]{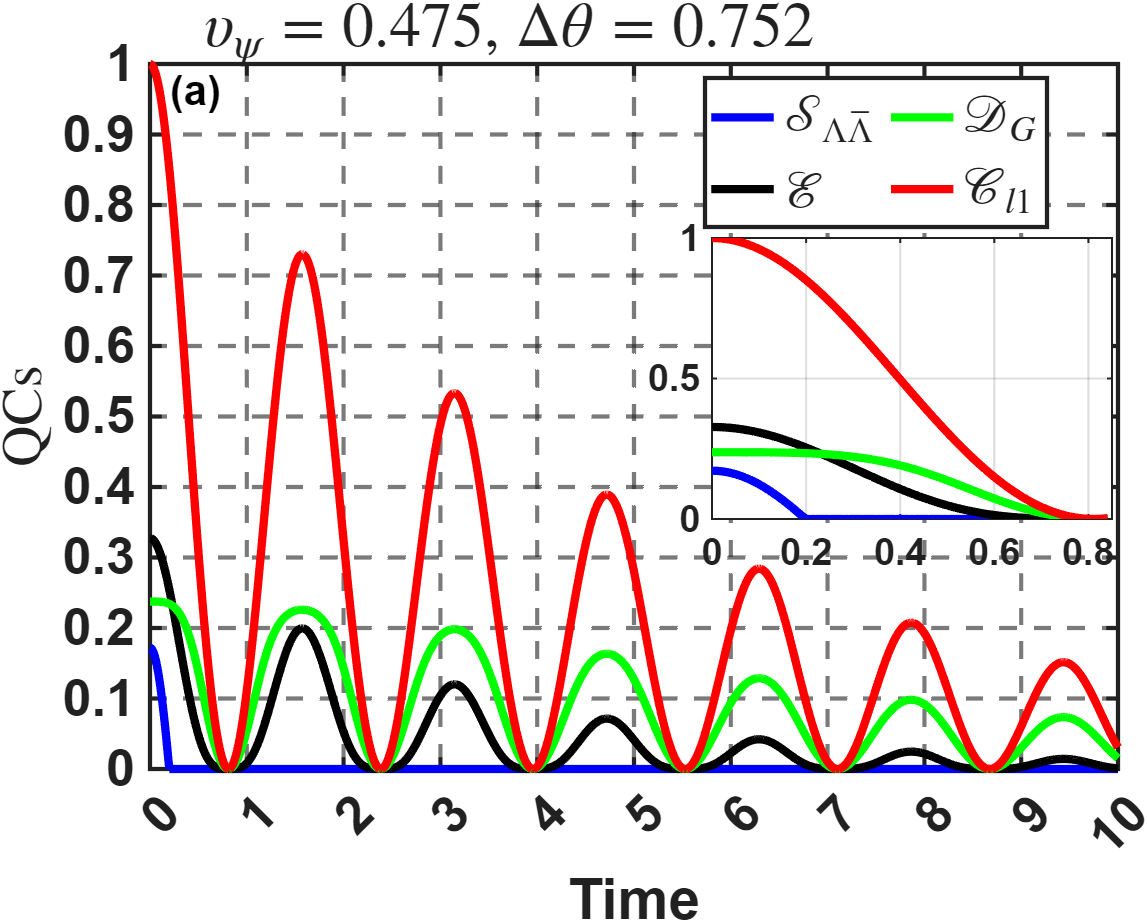}
    \includegraphics[scale=0.4]{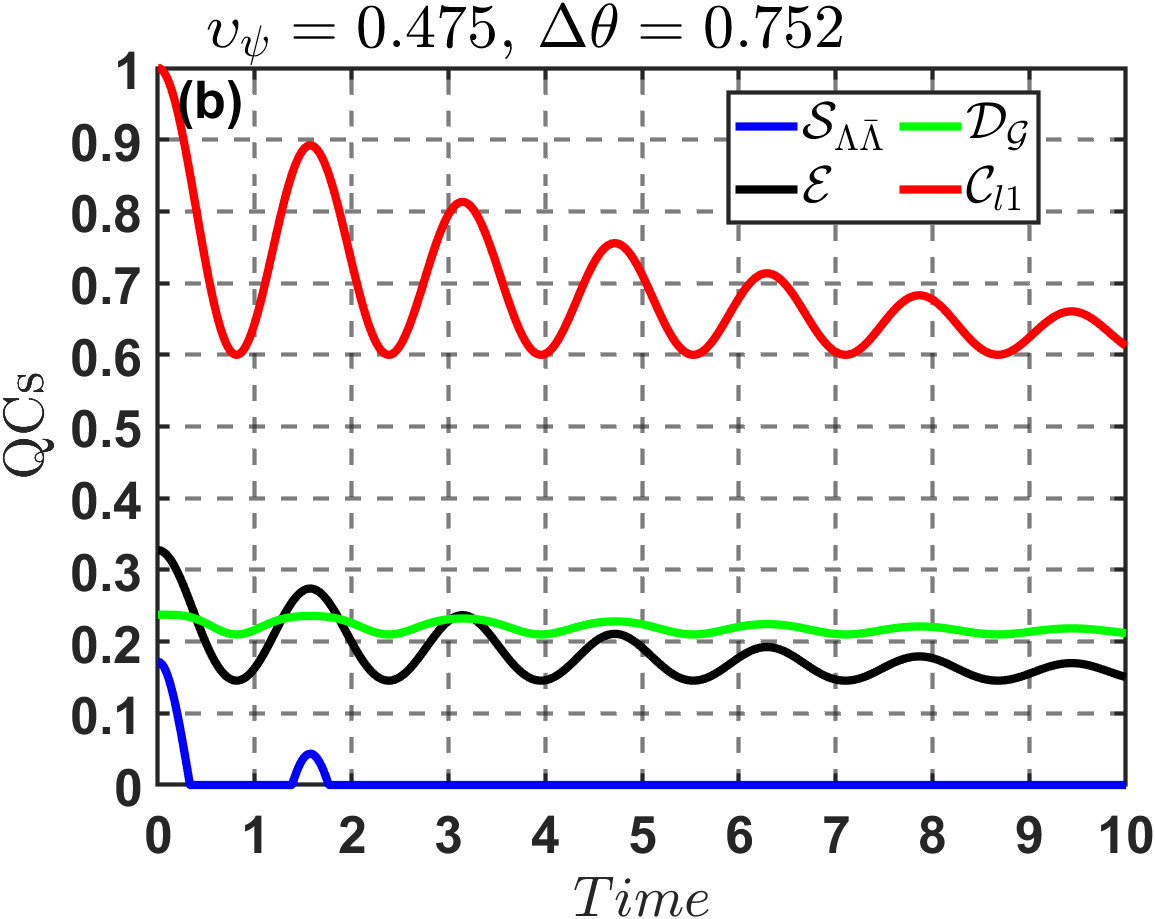}\\
    
    \includegraphics[scale=0.4]{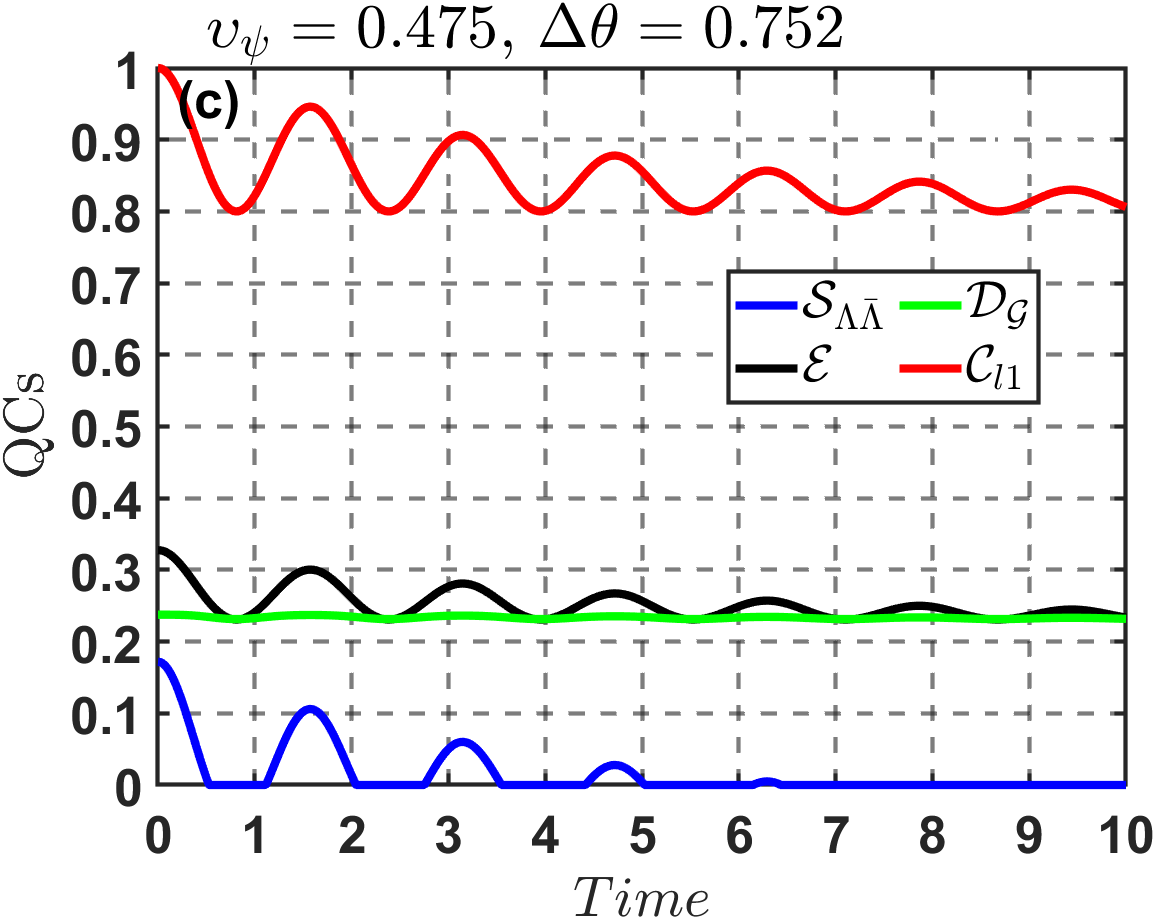}
    \includegraphics[scale=0.4]{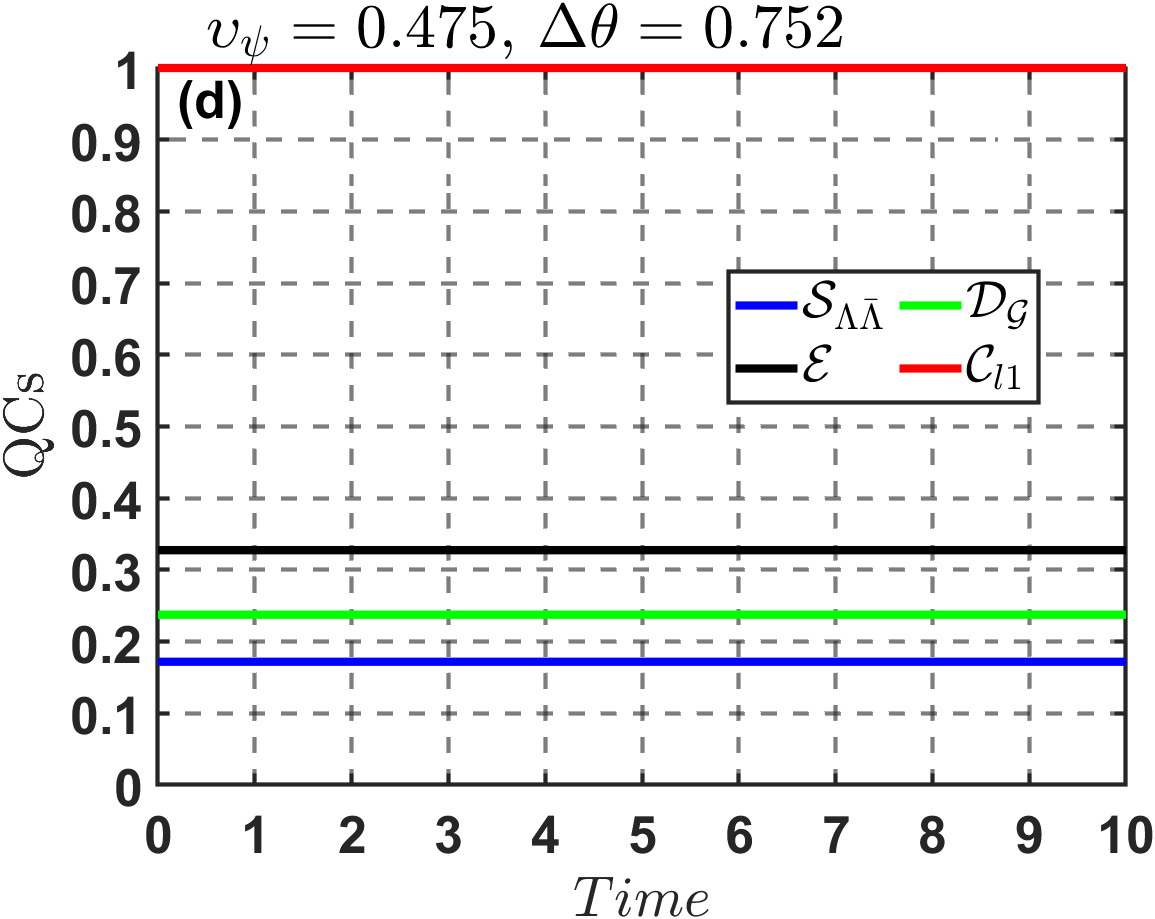}
\caption{Dynamical evolution of quantum steering $\mathcal{S}_{\Lambda\bar{\Lambda}}$, entanglement of formation \(\mathcal{E}\), GQD $\mathcal{D_{G}}$, and $l_1$-norm of quantum coherence $\mathcal{C}_{l1}$ in the non-Markovian regime for different values of $\mu$ : $\mu=0$ (a), $\mu=0.6$ (b), $\mu=0.8$ (c) and $\mu=1$ (d). With $\tau = 5$ for $\varphi = \pi/2$.}
\label{fig:NM}
\end{figure}

The observation that classical correlations enhance quantum steering, entanglement of formation, geometric quantum discord, and quantum coherence in hyperon-antihyperon states is of paramount importance. As depicted in Fig. \ref{fig:M}(d), when the classical correlations reach their maximum value of $\mu=1$, all quantum resources achieve its steady-state over time, signifying their robustness against decoherence. When $\mu<1$, the quantum resources degrade quickly over time, and the rate of degradation is inversely related to $\mu$. This decay emphasizes the critical role that classical correlations play in maintaining quantum properties. 

For $\mu = 0$, as illustrated in Fig.~\ref{fig:M}(a), the quantum resources are initially nonzero over the interval $Time \in [0, 0.3]$. Between $Time \in [0.3, 4]$, the $\Lambda\bar{\Lambda}$ qubits retain their entanglement even though the steerability measure vanishes ($\mathcal{S}_{\Lambda\bar{\Lambda}} = 0$) in $Time=0.3$, highlighting that entanglement can persist without steerability. This decoupling between the two resources is clearly visible by comparing Fig.~\ref{fig:h1} and Fig.~\ref{fig:e1}. Once entanglement disappears at $Time=4$ ($\mathcal{E} = 0$), GQD and quantum coherence continue to exist and decay monotonically. Notably, even after the disappearance of entanglement of formation, GQD and quantum coherence persist and tend toward an asymptotic regime, remaining at nearly constant values.Quantum coherence is the most general resource, serving as an upper bound for quantum discord, entanglement of formation, and quantum steering, which represent progressively more specific and fragile forms of correlations, as illustrated in Fig.~\ref{fig:M}(a–d).

In the non-Markovian regime, particularly for $\mu < 1$ [Fig.~\ref{fig:NM}(a-c)], the presence of memory effects leads to oscillatory behavior in quantum resources, particularly when classical correlations are diminished. This oscillatory behavior is the primary distinction between the Markovian and non-Markovian regimes. These correlations evolve in a sinusoidal and in-phase manner, highlighting the significant influence of non-Markovian memory effects on their dynamics. The amplitude of these oscillations decreases over time, reflecting a gradual decay of quantum resources. Notably, when classical correlations are absent ($\mu=0$), as depicted in Fig. \ref{fig:NM}(a), the quantum resources degrade quickly to reach its steady-state similar to those in the Markovian regime. When the classical correlations reach their maximum value (i.e. $\mu=1$), all quantum resources converge to a steady state over time, thereby illustrating their robustness against decoherence, as shown in Fig.~\ref{fig:NM}(d).

The behavior of quantum resources such as steerability, entanglement of formation, GQD, and quantum coherence, remains stable in both Markovian and non-Markovian regimes. However, the dynamics of these resources differ significantly depending on the regime. In the Markovian regime, the future states of the system are memoryless, whereas in the non-Markovian regime, past states influence the system's evolution. As illustrated in Fig.~\ref{fig:NM}(a), the resources are initially present in the interval \(Time \in [0, 0.2]\), and steerability rapidly vanishes around \(Time \approx 0.2\), as shown in Fig.~\ref{fig:h2}, while entanglement, geometric quantum discord (GQD), and quantum coherence maintain significant values. This result highlights the ability of certain quantum resources, such as entanglement, discord, and coherence, to persist even in the absence of steerability. In the interval \(Time \in [0.2, 0.7]\), a notable decrease in the entanglement of formation is observed, which may even temporarily vanish. In contrast, geometric discord and quantum coherence remain clearly non-zero. This observation emphasizes the increased robustness of these two resources, which can survive independently of the presence of entanglement, thereby revealing a hierarchy among the different forms of quantum correlations in a non-Markovian environment.

The hierarchy of quantum resources illustrates a structured progression of correlations, starting with quantum coherence, which constitutes the fundamental resource underlying all others. Quantum coherence bounds the geometric quantum discord (GQD), ensuring that GQD cannot exceed the coherence present in the system~\cite{RevModPhys1}. The GQD encompasses a broader range of correlations than entanglement, which represents a specific form of bipartite correlation related to the non-separability of states~\cite{RevModPhys2,RevModPhys3}. This hierarchical organization implies that the emergence of each new quantum property depends on the existence of the previous one. At the top of this hierarchy lies quantum steering, highlighting its greater sensitivity to environmental decoherence~\cite{PhysRevLett}. Thus, the hierarchy can be summarized as follows:
\[\text{Quantum coherence} \supseteq \text{Quantum discord} \supseteq \text{Entanglement of formation} \supseteq \text{Quantum steering},\]
emphasizing the central role of these resources in the dynamics of quantum information and the development of advanced quantum technologies~\cite{PhysRevA}.

\section{Conclusion}\label{sec:5} 

In this work, the dynamics of quantum coherence and quantum correlations in the process of $e^{+}e^{-} \to J/\psi \to \Lambda\bar{\Lambda}$ at the BESIII experiment are investigated. Quantum coherence is quantified using the l1-norm. Quantum steering is employed to measure the amount of steerability between the qubits, while entanglement of formation serves as an entanglement witness. Geometric quantum discord quantifies general quantum correlations beyond entanglement. The comparison between the quantum resources is discussed. Our results reveal that the quantum coherence and correlations are maximized at specific scattering angle $\varphi$, particularly at $\varphi =\pi/2$. We have discussed that the classical correlations can delay the decay of quantum correlations and coherence using feasible experimental parameters. Additionally, we have discussed the impact of non-Markovian memory effects, which play a crucial role in the stability of quantum coherence and quantum correlations. Crucially, our analysis confirms the theoretical hierarchy of quantum resources: the set of steerable states is a subset of entangled states, which in turn is a subset of states with non-zero geometric quantum discord. Quantum coherence further generalizes this hierarchy, as all states with discord exhibit coherence, but not vice versa-consistent with the results depicted in Figs. \ref{fig:NM}(a) and \ref{fig:M}(a), i.e. quantum coherence $\supseteq$ geometric quantum discord $\supseteq$ entanglement of formation $\supseteq$ quantum steerability. Our results have significant implications for future research on hyperon-antihyperon interactions, offering deeper insights into both strong and weak interactions in particle physics. They also pave the way for future experimental and theoretical studies on the engineering of quantum coherence and correlations, with potential applications in quantum technologies.

\end{document}